\documentclass[11pt, leqno]{article}
\usepackage{amsmath, amsthm, amssymb}
\input amssym.def 
\input amssym.tex
\def\utw{\smash{\rlap{\lower5pt\hbox{$\sim$}}}}
\def\udtw{\smash{\rlap{\lower6pt\hbox{$\approx$}}}}

\def\bbbr{{\rm I\!R}} 
\def\bbbn{{\rm I\!N}} 

\def\bbbc{{\mathchoice {\setbox0=\hbox{$\displaystyle\rm C$}\hbox{\hbox
to0pt{\kern0.4\wd0\vrule height0.9\ht0\hss}\box0}}
{\setbox0=\hbox{$\textstyle\rm C$}\hbox{\hbox
to0pt{\kern0.4\wd0\vrule height0.9\ht0\hss}\box0}}
{\setbox0=\hbox{$\scriptstyle\rm C$}\hbox{\hbox
to0pt{\kern0.4\wd0\vrule height0.9\ht0\hss}\box0}}
{\setbox0=\hbox{$\scriptscriptstyle\rm C$}\hbox{\hbox
to0pt{\kern0.4\wd0\vrule height0.9\ht0\hss}\box0}}}}
\def\bbbe{{\mathchoice {\setbox0=\hbox{\smalletextfont e}\hbox{\raise
0.1\ht0\hbox to0pt{\kern0.4\wd0\vrule width0.3pt
height0.7\ht0\hss}\box0}}
{\setbox0=\hbox{\smalletextfont e}\hbox{\raise
0.1\ht0\hbox to0pt{\kern0.4\wd0\vrule width0.3pt
height0.7\ht0\hss}\box0}}
{\setbox0=\hbox{\smallescriptfont e}\hbox{\raise
0.1\ht0\hbox to0pt{\kern0.5\wd0\vrule width0.2pt
height0.7\ht0\hss}\box0}}
{\setbox0=\hbox{\smallescriptscriptfont e}\hbox{\raise
0.1\ht0\hbox to0pt{\kern0.4\wd0\vrule width0.2pt
height0.7\ht0\hss}\box0}}}}
\def\bbbq{{\mathchoice {\setbox0=\hbox{$\displaystyle\rm Q$}\hbox{\raise
0.15\ht0\hbox to0pt{\kern0.4\wd0\vrule height0.8\ht0\hss}\box0}}
{\setbox0=\hbox{$\textstyle\rm Q$}\hbox{\raise
0.15\ht0\hbox to0pt{\kern0.4\wd0\vrule height0.8\ht0\hss}\box0}}
{\setbox0=\hbox{$\scriptstyle\rm Q$}\hbox{\raise
0.15\ht0\hbox to0pt{\kern0.4\wd0\vrule height0.7\ht0\hss}\box0}}
{\setbox0=\hbox{$\scriptscriptstyle\rm Q$}\hbox{\raise
0.15\ht0\hbox to0pt{\kern0.4\wd0\vrule height0.7\ht0\hss}\box0}}}}
\def\bbbt{{\mathchoice {\setbox0=\hbox{$\displaystyle\rm
T$}\hbox{\hbox to0pt{\kern0.3\wd0\vrule height0.9\ht0\hss}\box0}}
{\setbox0=\hbox{$\textstyle\rm T$}\hbox{\hbox
to0pt{\kern0.3\wd0\vrule height0.9\ht0\hss}\box0}}
{\setbox0=\hbox{$\scriptstyle\rm T$}\hbox{\hbox
to0pt{\kern0.3\wd0\vrule height0.9\ht0\hss}\box0}}
{\setbox0=\hbox{$\scriptscriptstyle\rm T$}\hbox{\hbox
to0pt{\kern0.3\wd0\vrule height0.9\ht0\hss}\box0}}}}
\def\bbbs{{\mathchoice
{\setbox0=\hbox{$\displaystyle     \rm S$}\hbox{\raise0.5\ht0\hbox
to0pt{\kern0.35\wd0\vrule height0.45\ht0\hss}\hbox
to0pt{\kern0.55\wd0\vrule height0.5\ht0\hss}\box0}}
{\setbox0=\hbox{$\textstyle        \rm S$}\hbox{\raise0.5\ht0\hbox
to0pt{\kern0.35\wd0\vrule height0.45\ht0\hss}\hbox
to0pt{\kern0.55\wd0\vrule height0.5\ht0\hss}\box0}}
{\setbox0=\hbox{$\scriptstyle      \rm S$}\hbox{\raise0.5\ht0\hbox
to0pt{\kern0.35\wd0\vrule height0.45\ht0\hss}\raise0.05\ht0\hbox
to0pt{\kern0.5\wd0\vrule height0.45\ht0\hss}\box0}}
{\setbox0=\hbox{$\scriptscriptstyle\rm S$}\hbox{\raise0.5\ht0\hbox
to0pt{\kern0.4\wd0\vrule height0.45\ht0\hss}\raise0.05\ht0\hbox
to0pt{\kern0.55\wd0\vrule height0.45\ht0\hss}\box0}}}}
\def\bbbz{{\mathchoice {\hbox{$\sf\textstyle Z\kern-0.4em Z$}}
{\hbox{$\sf\textstyle Z\kern-0.4em Z$}}
{\hbox{$\sf\scriptstyle Z\kern-0.3em Z$}}
{\hbox{$\sf\scriptscriptstyle Z\kern-0.2em Z$}}}}
\def\diameter{{\ifmmode\oslash\else$\oslash$\fi}} 
\def\init{\setcounter{equation}{0}}

\newtheorem{theoreme}{Theorem }[section]
\newtheorem{proposition}[theoreme]{Proposition}
\newtheorem{lemma}[theoreme]{Lemma}

\def\rr{\bbbr}

\def\nn{\bbbn}

\def\12{\frac{1}{2}}

\def\slim{\hbox{\rm s-}\lim}

\def\bull{$\sqcup \kern -0.645em \sqcap$}

\def\EG{{^{\scriptscriptstyle E} \kern -.05cmG}}

\def\c{{\rm c}}

\def\pfi2{P(\varphi)_{2}}

\newcommand{\beq}{\begin{equation}}
\newcommand{\eeq}{\end{equation}}
\newcommand{\bet}{\begin{theoreme}}
\newcommand{\eet}{\end{theoreme}}
\newcommand{\bel}{\begin{lemma}}
\newcommand{\eel}{\end{lemma}}
\newcommand{\bep}{\begin{proposition}}
\newcommand{\eep}{\end{proposition}}

\newcommand{\bear}[1]{\begin{array}{#1}}
\newcommand{\ear}{\end{array}}

\def\maath{\mathsurround=0pt }
\def\eqalign#1{\null\, \vcenter{\openup1\jot \maath
\ialign{\strut\hfil$\displaystyle{##}$&$\displaystyle{{}##}$\hfil
\crcr#1\crcr}}\,}

\def\CMP{Commun.\ Math.\ Phys.}
\def\JMP{J.\ Math.\ Phys.}
\def\LMP{Lett.\ Math.\ Phys.}
\def\today{\number\day .\space\ifcase\month\or
January\or February\or March\or April\or May\or June\or
July\or August\or September\or October\or November\or December\fi, \number \year}
\newcount \theoremnumber
\def\cleartheoremnumber{\theoremnumber = 0 \relax}
\def\versuch #1 #2 {
\vskip -.1 cm
\global \advance \equationnumber by 1
            $$\displaylines{ \rlap{ #1 } \hfill #2  \hfill } $$ 
\vskip  .1cm
\noindent}

\newcount \equationnumber

\newcount \refnumber

\def\# #1  {\global 
            \advance \equationnumber by 1
            $$ #1 \eqno ({\the\equationnumber}) $$ }

\def\% #1 { \global
            \advance \equationnumber by 1
            $$ \displaylines{ #1 \hfill \llap ({\the\equationnumber}) \cr}$$} 

\def\& #1 { \global
            \advance \equationnumber by 1
            $$ \eqalignno{ #1 & ({\the\equationnumber}) \cr}$$}
\def\bull{$\sqcup \kern -0.645em \sqcap$}
\def\Def#1{\vskip .6cm \goodbreak \noindent
                                     {\bf Definition.} #1 \goodbreak \vskip.8cm}
\def\Rem#1{\vskip .4cm \goodbreak \noindent
                                     {\it Remark.} #1 \goodbreak \vskip.5cm }
\def\Rems#1{\vskip .4cm \goodbreak \noindent
                                     {\it Remarks.} #1 \goodbreak \vskip.5cm }

\def\Pr#1{\goodbreak \noindent {\it Proof.} #1 \hfill \bull  \goodbreak \vskip.5cm}

\def\*{\vskip 1.0cm}      

\newcount \headlinenumber

\newcount \headlinesubnumber
\def\clearheadlinesubnumber{\headlinesubnumber = 0 \relax}
\def\Hl #1 {\goodbreak
            \cleartheoremnumber
            \clearheadlinesubnumber
            \advance \headlinenumber by 1
            {\bf \noindent {\the\headlinenumber}. #1}
            \nobreak \vskip.4cm \rm \noindent}

\font\css=cmss10
\font\Rosch=cmr10 at 9.85pt
\font\Cosch=cmss12 at 9.5pt
\font\rosch=cmr10 at 7.00pt
\font\cosch=cmss12 at 7.00pt
\font\nosch=cmr10 at 7.00pt
\def\Z                 {\hbox{{\css Z}  \kern -1.1em {\css Z} \kern -.2em }}
\def\R                 {\hbox{\raise .03ex \hbox{\Rosch I} \kern -.55em {\rm R}}}
\def\N                 {\hbox{\rm I \kern -.55em N}}
\def\C                 {\hbox{\kern .20em \raise .03ex \hbox{\Cosch I} \kern -.80em {\rm C}}}

\def\r                 {\hbox{\raise .03ex \hbox{\rosch I} \kern -.45em \hbox{\rosch R}}}
\def\n                 {\hbox{\hbox{\rosch I} \kern -.45em \hbox{\nosch N}}}
\def\c                 {\hbox{\raise .03ex \hbox{\cosch I} \kern -.70em \hbox{\rosch C}}}

\def\z                 {\hbox{\kern 0.2em {\cal z}  \kern -0.6em {\cal z} \kern -0.3em  }}
\def\unit                  {\hbox{\rm \thinspace \thinspace \thinspace \thinspace
                                  \kern -.50em  l \kern -.85em 1}}
\def\Tr                {{\rm Tr}}
\def\A                 {{\cal A}}

\def\B                 {{\cal B}} 
\def\M                 {{\cal M}}
\def\H                 {{\cal H}} 
 
\def\O                 {{\cal O}}

\def\slim              {\mathop{\rm s-lim}}
\begin{document}
%
%
\title{The Relation between KMS States for Different Temperatures\protect\footnotetext{AMS 1991 {\it{Subject 
Classification}}. 81T05} \protect}          
%
%
\author{Christian D.\ J\"akel\footnote{
christian.jaekel@mathematik.uni-muenchen.de, Math.\ Inst.\ d.\ LMU,
Theresienstr.~39, 80333 M\"unchen; 
partially supported by the IQN network of the DAAD and the IHP network HPRN-CT-2002-00277
of the European Union.
}}        
%
%
\date{August 2003}
%
%
\maketitle
\abstract{Given a thermal field theory for some temperature $\beta^{-1}$, 
we construct the theory at an arbitrary temperature $ 1 / \beta'$.
Our work is based on a construction invented by Buchholz and Junglas, which 
we adapt to thermal field theories. 
In a first step we construct states which closely resemble
KMS states for the new temperature in a 
local region $\O_\circ \subset \rr^4$, but coincide with the given
KMS state in the space-like complement of a slightly larger region~$\hat{\O}$.
By a weak*-compactness argument there always exists a convergent subnet of
states as the size of~$ \O_\circ$ and $ \hat{\O}$ tends towards $ \rr^4$.
Whether or not such a limit state is a global KMS state for the new temperature, 
depends on the surface energy contained 
in the layer in between the boundaries of $ \O_\circ$ and 
$ \hat{\O}$. 
We show that this surface energy can be controlled by a generalized cluster condition.}

\tableofcontents

\section{Introduction}
\init\label{introd}
%
%
%
%
%
%
%
\noindent
A quantum field theory 
can be specified by a $C^*$-algebra $\A$ together with a
net
\[  \O \to \A (\O), \qquad \O \subset \R^4,  \]
of subalgebras associated with open, bounded space-time regions $\O$ in Minkowski space
(as described in the monograph by Haag \cite{H}; see also \cite{HK}).
The Hermitian elements of~$\A(\O)$ 
are interpreted as the observables which
can be measured at times and locations in~$\O$.  
Technically the algebra $\A(\O)$ may be thought of as being generated by bounded functions of the underlying smeared quantum
fields (see, e.g., \cite{BoY}). For instance,
if $\phi(x)$ is a hermitian quantum field and if $f(x)$ is a real test 
function with support in a bounded region~$\O$ of space-time, then the unitary
operator 
$ a := \exp \bigl( i \int {\rm d}x \, f (x) \phi (x) \bigr) $
is a typical element of~$\A (\O)$. In this way the quantum fields provide 
a ``coordinate system'' for the algebra~$\A$. However, as emphasized by Haag and Kastler,
only the algebraic relations between the elements of~$\A$ are of physical significance. 

If the time evolution is given by a strongly continuous one-parameter group of 
auto\-morphisms $\{ \tau_t \}_{t \in \r}$ of $\A$, 
then the pair $(\A, \tau)$ forms a $C^*$-dynamical system.
Such a description of a QFT fits nicely into the structure of algebraic quantum statistical mechanics
(see, e.g., \cite{BR}\cite{E}\cite{R}\cite{Se}\cite{Th}) and we can therefore rely on this well-developed framework. 

Up till now
non-relativistic quantum field theories and spin systems were favoured in the
framework of algebraic quantum statistical mechanics. 
In low dimensions the latter have been worked out in 
great detail (see, e.g., \cite{BR}). Only recently the benefits of formulating 
thermal field theory  in the algebraic framework  
were emphasized in a series of papers~\cite{BJu 86}\cite{BJu 89}\cite{BB 94}\cite{N}\cite{Jae 98}\cite{Jae 99}\cite{Jae d}.

Equilibrium states can be characterized by first principles in the algebraic framework: equilibrium states are invariant under
the time-evolution  and stable
against small dynamical (or adiabatic \cite{NT}) perturbations 
of the time-evolution~\cite{HKTP}. 
Adding a few technical assumptions such a heuristical
characterization of an equilibrium state leads to a sharp mathematical criterion~\cite{HHW},
named for Kubo \cite{K}, Martin and Schwinger~\cite{MS}: 

\Def{A state
$\omega_\beta$ over $\A$ is called a $(\tau , \beta )$-KMS state for 
some $\beta \in \R \cup \{ \pm \infty \}$, if 
\beq
\omega_\beta \bigl( a \tau_{i \beta} (b) \bigr) = \omega_\beta (b a)
\label{KMS}  \eeq
for all $a, b$ in a norm dense,
$\tau$-invariant $*$-subalgebra of $\A_\tau$. Here  $\A_\tau  \subset \A$
denotes the set of analytic elements for $\tau$.}

We note that  there are $C^*$-dynamical 
systems~$(\A, \tau)$, for which  a KMS state exists at one and only one 
value~$\beta \in \R$  (see \cite[5.3.27]{BR}). But for a QFT on can
specify conditions on the phase-space properties  in the vacuum representation, 
such that KMS states exist for all temperatures $\beta^{-1} > 0$ \cite{BJu 89}.  
These conditions exclude (see  \cite{BJu 86}) the class of models with a countable 
number of free scalar particles proposed by Hagedorn \cite{Ha}. These models obey all the Wightman and 
Haag-Kastler axioms but they do not allow equilibrium states above a certain critical 
temperature.  

For a generic model one expects that for high temperatures
and low densities the set of KMS states contains a unique element\footnote
{\rm For non-relativistic fermions with pair-interaction see 
\cite{Jae 95}.},
whereas at low temperature it should 
contain many disjoint extremal KMS states and their 
convex combinations corresponding to various thermodynamic phases and their possible mixtures.
The symmetry, or lack of symmetry of the extremal
KMS states is automatically determined by this decomposition. Consequently, spontaneous symmetry breaking may occur, when we change the temperature in the sequel.
\vskip .2cm

Given a KMS state $\omega_\beta$ over $\A$
the GNS-representation $(\pi_\beta, \H_\beta, \Omega_\beta)$ provides a net of von Neumann algebras:
\[ \O \to {\cal R}_\beta (\O) := \pi_\beta \bigl( \A(\O) \bigr)'' , \qquad \O \in \R^4. \]
Under fairly general circumstances KMS states for different 
values of the temperature~$\beta^{-1}$ lead to unitarily inequivalent 
GNS-representations (see \cite{T}\cite[5.3.35]{BR}).
Hence thermal field theories for different temperatures are frequently treated as completely 
disjoint objects even if they refer to the same vacuum theory, i.e., even if
they show identical interactions on the microscopic level. To 
understand the relations between these `disjoint thermal field theories'
seems to be highly desirable. 

One simple case is well known  (\cite[8.12.10]{Pe}):
Assume that the time-evolution~$ \tau$ can be approximated by a net of inner automorphisms 
such that, for
$a \in \A$ fixed,  
\[  \lim_{\Lambda \to \infty} \| \tau_z (a) - 
{\rm e}^{i z  h_\Lambda} a {\rm e}^{-i z  h_\Lambda} \| = 0 , 
\qquad  h_\Lambda = h_\Lambda^* \in \A , \]
uniformly in $z$ on compact subsets of~$\C$. If $ (\A, \tau)$  
has a KMS state $\omega_{\beta}$ at some~$ \beta \ne 0$,  
then the net of states $\Lambda \mapsto \omega_\Lambda$, 
\[ \omega_\Lambda (a) = \frac{ \omega_\beta 
\bigl( {\rm e}^{ \frac{1}{2}(\beta - \beta')  h_\Lambda} 
a {\rm e}^{ \frac{1}{2}(\beta - \beta') h_\Lambda} \bigr) 
}{\omega_{\beta} \bigl( {\rm e}^{ (\beta - \beta') 
h_\Lambda} \bigr) },
\qquad a \in \A  , \] 
has convergent subnets and the limit points 
$ \omega_{\beta'}  := \lim_{\Lambda} \omega_{\Lambda}$ 
are $(\tau, \beta')$-KMS states for the new temperature $1/\beta'$
$ (0 < \beta' < \infty)$. 

But in general, phase transitions may occur while 
we change the temperature. Consequently ``... there is no simple prescription for connecting 
the~$(\tau, \beta)$--KMS states for different~$\beta$'s'' (c.f.\ \cite{BR} p.78). 
Nevertheless, we will provide a prescription which covers, as
far as relativistic systems are concerned, the physically relevant cases. 

We start form a thermal field
theory $\O \to {\cal R}_\beta (\O)$, whose number of local degrees of freedom 
is restricted in a physically sensible manner.
Using a method, which is essentially due to Buchholz and Junglas \cite{BJu 89}, 
we construct a KMS state $\omega_{\beta'}$ and a thermal field theory 
\[ \O \to {\cal R}_{\beta'} (\O), \qquad \beta' \in \R^+, \]
for a new temperature $1 / \beta' > 0$.
Although we almost repeat their line of arguments, there are some 
nontrivial deviations due to the mathematical structure we encounter in thermal field theory. 

In a first step we construct product states $\omega_\Lambda$, $\Lambda = (\O_\circ, \hat{\O})$,
which---up to boundary effects---resemble KMS states 
for the new temperature $1 / \beta'$ in a local region $\O_\circ \subset \R^4$, but 
coincide with the given KMS state~$\omega_\beta$
in the space-like complement of a slightly larger region~$\hat{\O}$:
\[ \omega_\Lambda (ab)  = \omega_\Lambda (a) \cdot \omega_\beta (b) \qquad  
\forall a \in \A (\O_\circ), \quad \forall b \in \A(\hat {\O}'). \]
At this point our
method is semi-constructive; the product states~$\omega_\Lambda$
is not uniquely fixed.
Intuitively the choice of a particular product state $\omega_\Lambda$ corresponds to a 
choice of the boundary conditions which decouple the local region $\O_\circ$, 
where the state already resembles an equilibrium state for the
new temperature, from the space-like complement of~$\hat{\O}$. 
Different choices $\omega_\Lambda$, ${\omega_\Lambda}'$ should manifest themselves 
in different expectation values
for observables localized in between the two regions $\O_\circ$ and $\hat{\O}'$. 
I.e., we expect
\[ \omega_\Lambda  \ne {\omega_\Lambda}' \quad 
\Rightarrow \quad  \exists a \in \A (\O_\circ' \cap \hat {\O}) \hbox { such that } 
\omega_\Lambda (a) \ne {\omega_\Lambda}' (a). \]

It follows from standard compactness arguments that the
net of states $\Lambda \to \omega_\Lambda$ has convergent subnets.
Whether or not these subnets converge to a global KMS states for the new temperature
depends on the surface energy contained in between the two 
regions~$\O_\circ$ and~$\hat{\O}'$ as their size increases. 
Introducing an auxiliary structure, which can be understood as a local
purification, and assuming a  cluster condition, we will 
control these surface energies in all thermal
theories which satisfy a certain ``nuclearity condition'' (see, e.g., \cite{BW}\cite{BD'AL 90a}\cite{BD'AL 90b}\cite{BY}
for related work). 
Consequently, we can single out (generalized) sequences $\Lambda_i = \bigl(\O_\circ^{(i)}, \hat{\O}^{(i)}\bigr)$ 
such that the limit points\footnote
{\rm We have simplified the notation here. 
In fact, we will have to adjust the relative sizes of a  
triple $
\Lambda_i = (\O_\circ^{(i)},\O^{(i)}, \hat{\O}^{(i)})$ of space-time regions.} 
\[  \omega_{\beta'} (a) := \lim_{i \to \infty} \omega_{\Lambda_i} (a) ,  
\qquad a \in \A, \]  
are KMS states for the new  temperature $1/ \beta'$
$ (0 \le \beta' \le \infty)$. We emphasize that phase transitions are not excluded by our method: 
by choosing different ``boundary conditions''
we may encounter disjoint KMS states for the new temperature in the thermodynamic limit.

Loosely speaking, we provide a method to heat up or cool down 
a quantum field theory. 

\vskip 1cm

\section{Definitions and preliminary results}

For the Lagrangian formulation of a thermal field theory
we refer the reader to the books by Kapusta \cite{Ka}, Le Bellac~\cite{L} and Umezawa \cite{U}, and 
the excellent review article 
by Landsman and van Weert~\cite{LvW}. Recent work in the Wightman framework can be found 
in~\cite{BB 92}\cite{BB 95}\cite{BB 96}\cite{St}. In this section we will outline the basic structure of a
thermal field theory in the algebraic framework.

\subsection{List of assumptions}

\noindent
Although it would be more natural---from the viewpoint of algebraic quantum statistical mechanics---to 
start from a $C^*$-dynamical system $(\A, \tau)$ and then characterize equilibrium 
states $\omega_\beta$ and thermal representations $\pi_\beta$
with respect to the dynamics, we will assume here that we are given a
thermal field theory $\O \to {\cal R}_\beta (\O)$ acting on some Hilbert space $\H_\beta$. 
How we can reconstruct a $C^*$-dynamical system $(\A, \tau)$ 
from the $W^*$-dynamical system $({\cal R}_\beta, \hat \tau)$ is well known and will be indicated 
in the next subsection ($\hat \tau$ will be defined in (\ref{tauhat})). 

\vskip.5cm
We now provide a list of assumptions:
\vskip.3cm

\noindent
i.) (Thermal QFT). A thermal QFT is specified by a von Neumann algebra ${\cal R}_\beta$,
acting on a separable Hilbert space~$\H_\beta$, together with a
net
\versuch {(Net structure)} { \O \to {\cal R}_\beta (\O), \qquad \O \subset \R^4, }
of subalgebras associated with open bounded space-time regions $\O$ in Minkowski space.
The net $\O \to {\cal R}_\beta (\O)$ satisfies
\versuch {(Isotony)} {{\cal R}_\beta (\O_1) \subset {\cal R}_\beta (\O_2) 
\quad \hbox{if} \quad \O_1 \subset \O_2  }
and
\versuch {(Locality)} {{\cal R}_\beta (\O_1) \subset {\cal R}_\beta (\O_2)' 
\quad \hbox{if} \quad  \O_1 \subset \O_2' .}
As before, $\O'$ denotes the space-like complement of $\O$.
\goodbreak

\vskip .3cm
\noindent
ii.) (Dynamical law). The time-evolution $\hat \tau \colon t \mapsto \hat \tau_t$,
\[  \label{tauhat} \hat  \tau_t (\, . \,) = {\rm e}^{iH_\beta t } \, \,  .\,  \,{\rm e}^{- iH_\beta t } ,\]
is induced by a strongly continuous one-parameter group of unitaries 
$\bigl\{Ê{\rm e}^{i H_\beta t} \bigr\}_{ t \in \rr}$. 
It acts geometrically, i.e.,
$\hat \tau_t \bigl( {\cal R}_\beta (\O) \bigr)
\subset {\cal R}_\beta (\O + t e ) $ for all $t \in \R$.  
Here $e$ is the unit-vector in the time-direction in the Lorentz-frame distinguished by the KMS state. 

\vskip .3cm
\noindent
iii.) (Unique KMS vector). 
There exists a distinguished vector
$\Omega_\beta$, cyclic and separating for~${\cal R}_\beta$, such that the associated 
vector state $\omega_\beta (\, . \, ):= (\Omega_\beta \, , \, . \, \, \Omega_\beta)$ 
satisfies the KMS condition~(\ref{KMS})~w.r.t.\ the time-evolution $\tau$.
Restricting attention to pure phases we assume that~$\Omega_\beta$ is the 
unique---up to a phase---normalized eigenvector with eigenvalue $\{ 0 \} $ of $H_\beta$.

\vskip .3cm
\noindent
iv.) (Reeh-Schlieder property). The KMS vector
$\Omega_\beta$ is cyclic and separating for the local algebra ${\cal R}_\beta (\O)$,
if the space-like complement of $\O \subset \R^4$ is not empty. 

\vskip .3cm
\noindent
v.) (Nuclearity condition). The thermal field theory $ \O \to
{\cal R}_\beta (\O )$ has the following phase-space properties:
for $\O$ bounded the maps $\Theta_{\alpha, {\cal O}}
\colon {\cal R}_\beta (\O) \to \H_\beta $ given by 
\[  \Theta_{\alpha, {\cal O}} (A) = {\rm e}^{- \alpha \beta H_\beta } A \Omega_\beta , 
\qquad 0 \le \alpha \le 1/2, \]
are nuclear for $0 < \alpha < 1/ 2$ and the nuclear norm 
(for $\alpha \searrow 0$ or $\alpha \nearrow 1/2$ and large diameters~$r$
of~$\O$) satisfies
\beq \| \Theta_{\alpha, {\cal O}} \|  \le {\rm e}^{cr^d \bigl 
( \alpha^{-m} + (1 / 2 - \alpha)^{-m} \bigr) }, \label{nuc} \eeq
where $c, m, d$ are positive constants. (We expect that the constant $d$ in this bound
can be put equal to the dimension of space in realistic theories, but we do not make 
such an assumption here. The constant $m> 0$ may depend on  
the interaction and the KMS state.)

\vskip .3cm
\noindent
vi.) (Regularity from the outside). The net $\O \to {\cal R}_\beta (\O)$ 
is regular from the outside, i.e.,
\[ \bigcap_{\hat{\cal O}^{(i)} \supset \O} 
{\cal R}_\beta \bigl( \hat{\O}^{(i)} \bigr) = {\cal R}_\beta (\O) , 
\qquad \hat{\O}^{(i)} \searrow \O. \]
(This property can usually be achieved by defining the local algebras in an appropriate way.)
\vskip .3cm
\noindent
vii.) (Cluster assumption). Let $\O_\circ$ and $\O$ be two space-time regions 
such that
$ \O_\circ + te \subset \O$  for~$|t| < \delta_\circ$. 
Let $J$ denote the modular conjugation (see Subsection 2.3) for the pair $({\cal R}_\beta, \Omega_\beta)$.
Let 
$M_j \in {\cal R}_\beta \bigl( \O_\circ \bigr) \vee J {\cal R}_\beta \bigl( \O_\circ \bigr)  J$ and 
$N_j \in \bigl({\cal R}_\beta \bigl( \O\bigr) \vee J {\cal R}_\beta \bigl( \O\bigr) J \bigr)'$. 
Then, for $\delta_\circ$ large compared to the thermal wave-length $\beta$,
\beq \Bigl| \sum_{j=1}^{N}( \Omega_\beta  \, , \,  M_j \Omega_\beta )(\Omega_\beta \, , \, N_j\Omega_\beta )
- ( \Omega_\beta \, , \,  M_j N_j\Omega_\beta )
\Bigr|  
\le
c' \, r_\circ^{d'} \delta_\circ^{-{\gamma} }
\cdot \Bigl\| \sum_{j=1}^{N} M_j N_j \Bigr\|,  \label{cluster} \eeq
where $c', d'$ and $\gamma$ are positive constants which do not depend on $ \O_\circ$ or $ \O$. 
Here $r_\circ$ denotes the diameter of $\O_\circ$.

\Rems{\vskip.2cm
\noindent
i.) The Reeh-Schlieder property is a consequence \cite{Jae 00} of 
additivity\footnote
{\rm The net $ \O \to {\cal R}_\beta (\O)$ is called additive  if
$ \cup_i \O_i = \O \Rightarrow \vee_i {\cal R}_\beta (\O_i) 
= {\cal R}_\beta (\O)  $. Here $
 \vee_i {\cal R}_\beta (\O_i)$
denotes the von Neumann algebra generated by the algebras 
$ {\cal R}_\beta (\O_i)$.} 
and the relativistic KMS condition
proposed by Bros and Buchholz \cite{BB 94}. If the KMS state is locally normal  w.r.t.\ the vacuum 
representation, then the standard KMS condition 
(together with additivity of the net in the vacuum representation) is sufficient to 
ensure the Reeh-Schlieder property of the KMS vector~$\Omega_\beta$~\cite{J}. 
\vskip.2cm
\noindent
ii.) If the KMS state is locally normal w.r.t.\ the vacuum representation, then
it is sufficient to assume that 
\[ \bigcap_{\hat{\cal O}^{(i)} \supset \O} {\cal R} \bigl( \hat{\cal O}^{(i)} \bigr) 
= {\cal R} (\O) , 
\qquad \hat{\O}^{(i)} \searrow \O, \]
holds true in the vacuum representation. For the free scalar field this property was
shown by Araki~\cite{A 64}.
\vskip.2cm
\noindent
iii.) One might try to establish the cluster condition starting from a sharper
nuclearity condition. For instance, we might assume that 
the map $\Theta^\sharp_{\alpha, {\cal O}} \colon {\cal R}_\beta (\O) \to \H_\beta$
given by
\[ \Theta^\sharp_{\alpha, {\cal O}} (A) = {\rm e}^{- \alpha | H_\beta |} \bigl( A 
- (\Omega_\beta \, , \, A \Omega_\beta) \bigr) \Omega_\beta,
\qquad \alpha > 0, \] 
is nuclear too and satisfies (for $\alpha^m$ large in comparison with $ r^d $) 
the following bound on its nuclear norm
\[  \| \Theta^\sharp_{\alpha, {\cal O}} \|  \le c'  \cdot r^d \alpha^{-m} . \]
Formally the bound on the nuclear norm $\| \Theta^\sharp_{\alpha, {\cal O}} \|$  follows
from taking the limit $\alpha^{ m}$ large in comparison with $r^d$ in the expression  
$\exp (c  r^d \alpha^{-m}) - 1 $, 
where the one is due to the subtraction of the thermal expectation value. (The expression 
$\exp (c  r^d \alpha^{-m})$ should provide an upper bound for the nuclear norm of the map $A \mapsto \exp( - \alpha | H_\beta |) A \Omega_\beta$, where $\alpha>0$.)
\vskip.2cm
\noindent
iv.) The product state appearing in (\ref{cluster}) is induced by a  product vectors~$\chi_i $,
which satisfies
\[ ( \chi_i \, , \, M N  \, \chi_i)  = (\Omega_\beta \, , \, M \Omega_\beta)
(\Omega_\beta \, , \, N \Omega_\beta) 
\]
for $M \in {\cal M}_\beta \bigl( \O_\circ^{(i)} \bigr)$ and $N \in {\cal M}_\beta \bigl( \O^{(i)} \bigr)' $.
The convergence of the product vector $\chi \to \Omega_\beta$ follows from $\bigr(\cup_\O {\cal M}_\beta ( \O) \bigl)'= \unit $ (see \cite{DADFL}).}

\vskip 1cm
\subsection{The restricted $C^*$-dynamical system}

\noindent
If the weakly continuous one-parameter group $\hat \tau \colon t \mapsto \hat \tau_t$
fails to be strongly continuous, then we can reconstruct the underlying $C^*$-dynamical system
by  a suitable smoothening procedure (once again we refer to \cite[1.18]{S}): 
given a thermal field theory $\O \to {\cal R}_\beta (\O)$ there exists 
\vskip .3cm
\halign{ \indent \indent \indent #  \hfil & \vtop { \parindent = 0pt \hsize=34.8em
                            \strut # \strut} 
\cr 
(i)    & a $C^*$-algebra $\A$ and a representation $\pi_\beta \colon
\A \to \B(\H_\beta)$ such that $\pi_\beta (\A)$ is a 
$\sigma$-weakly dense $C^*$-subalgebra of ${\cal R}_\beta$;
\cr
(ii)    & a net $\O \to \A(\O)$ of $C^*$-subalgebras of $\A$ 
such that $\pi_\beta \bigl(\A (\O)\bigr)$ is a 
$\sigma$-weakly dense $C^*$-subalgebra of ${\cal R}_\beta(\O)$ for all $\O \subset \R^4$;
\cr
(iii)    & a strongly continuous automorphism group $t \mapsto \tau_t$ of $\A$  
such that $\pi_\beta \bigl( \tau_t (a) \bigr) = \hat \tau_{t } ( \pi_\beta ( a) \bigr)$
for all $a \in \A$. 
\cr}
\vskip .5cm
\noindent
Moreover, the net $\O \to \A(\O)$ satisfies isotony and locality and $\tau$
respects the local structure of the net $\O \to \A(\O)$, i.e.,
$\tau_t \bigl( \A (\O) \bigr) = \A (\O + te)$
for $t \in \rr$.
\vskip .5cm

We can now introduce subalgebras $\A_p$ of
almost local elements in $\A$ which are analytic with 
respect to time-translations \cite{BJu 89}. 
For the existence of these subalgebras
it is crucial that the time-evolution~$t \mapsto \tau_t$ is strongly continuous, i.e., if we fix some $a \in \A$, then
$\lim_{t \to 0}  \| \tau_t (a) - a \| = 0$.
 
\begin{lemma}
{{\rm (Buchholz and Junglas)}. Let $p \in \N$ be fixed and let $\A_p \subset \A$ be 
the $*$-algebra generated by all
finite sums and products of operators of the form
\[ a(f) = \int {\rm d}t \, f(t) \tau_t (a), \]
where $f$ is any one of the functions 
\[ f(t) = const. \, {\rm e}^{- \kappa(t+ w)^{2p}} \]
(with $\kappa > 0$, $w \in \C$) and $a \in \cup_{\cal O} \A(\O) $ is 
any strictly local operator. It follows that
\vskip .3cm
\halign{ \indent \indent \indent #  \hfil & \vtop { \parindent = 0pt \hsize=34.8em
                            \strut # \strut} 
\cr 
(i)    &  each $b \in \A_p$ is an analytic element with respect to $\tau$, 
i.e., the operator-valued function $t \mapsto \tau_t(b)$ can be 
extended to a holomorphic function on \C ;
\cr
(ii)    & each $b \in \A_p$ is almost local in the sense that for any $r^{(i)} > 0$ 
there is a local 
operator $b^{(i)} \in \A(\O^{(i)})$ such that
\[  \| b^{(i)} - b \| \leq C \, {\rm e}^{- \kappa( r^{(i)} /2)^{2p}}, \qquad \kappa >0, \]
where the constant $C > 0$ does not depend on $r^{(i)}$;
\cr
(iii)    & the algebra $\A_p$ is invariant under $\tau_z$, $z \in \C$, and norm dense in $\A$.
\cr} }
\end{lemma} 

\vskip .3cm

The new state $\omega_{\beta'}$, which we will construct in the sequel, will be a 
$(\tau, \beta')$-KMS state for the pair $(\A, \tau)$. More precisely, it
will satisfy the KMS condition (\ref{KMS}) for $a, b \in \A_p$ for some~$p$ ($p$ will be specified in Subsection 4.3). 
As we have just seen, $\A_p$ is a
norm dense, $\tau$-invariant subalgebra of~$\A_\tau$.

\Rem{If the new state $\omega_{\beta'}$  is locally normal w.r.t.\ $\pi_\beta$,  then one 
might expect that the KMS condition extends to
\[  {\cal F}:= \overline{ \bigcup_{\O \in \r^4} {\cal R}_\beta (\O) }^{C^*}. \]
However, the representations $\pi_{\beta'}$ and $\pi_\beta$ of ${\cal F}$ will be inequivalent for $\beta' \ne \beta$,  and therefore the weak closures $\pi_{\beta'} ({\cal F})''$ and $\pi_{\beta} ({\cal F})''$ will in general be non-isomorphic.}

\vskip 1cm

\subsection{The opposite net of local algebras}

\noindent
By assumption the KMS vector $\Omega_\beta$ is cyclic and separating for ${\cal R}_\beta$.
Thus Tomita-Takesaki theory applies: the polar decomposition 
%
%
of the closeable operator  
$S_\circ  \colon A \Omega_\beta \mapsto  A^* \Omega_\beta$, $A \in {\cal R}_\beta$, 
provides a conjugate-linear isometric mapping~$J$
from $\H_\beta$ onto $\H_\beta$ and a positive self-adjoint 
(in general, unbounded, but densely defined and invertible) operator $\Delta$ acting 
on~$\H_\beta$.  The modular conjugation $J$ satisfies $J^2 = \unit $ and
\[ J \Delta^{1/2} A \Omega_\beta = A^* \Omega_\beta 
\qquad
\forall A \in {\cal R}_\beta. \]
$\Delta$ is called the modular operator. $J$ induces a $*$-anti-isomorphism $j \colon 
A \mapsto JA^*J$ between the algebra of quasi-local observables ${\cal R}_\beta$ and its commutant
(Tomita's theorem). The opposite net
\[  \O \to j \bigl( {\cal R}_\beta (\O) \big), \qquad \O \subset \R^4, \]
provides a perfect mirror image of the net of local observables.
The unitary operators~$\Delta^{is}$, $s \in \R$, induce a one-parameter group of
$*$-automorphism 
$\sigma \colon s \mapsto \sigma_s$ of ${\cal R}_\beta$, 
\[ \sigma_s (A) = \Delta^{is} A \Delta^{-is} , \qquad s \in \R, \quad A \in {\cal R}_\beta . \]
$\sigma$ is called the modular automorphism. Takesaki has shown that 
$\omega_\beta$ is a $(\sigma, -1)$-KMS state. Moreover,
$\sigma$ is uniquely determined by this condition and consequently
$\Delta^{is} =\exp \bigl( -i s \beta  H_\beta \bigr)$. 

We conclude that in a thermal field theory
the modular automorphism $\sigma$ coincides---up to a scaling factor---with
the time-evolution~$\hat \tau$. Consequently, the modular automorphism respects the net structure too, i.e.,
\beq  \sigma_s \bigl( {\cal R}_\beta (\O) \bigr) = {\cal R}_\beta (\O + s \beta \cdot e) 
\qquad \forall s \in \R. \label{a}\eeq
The real parameter $\beta \in \R^+$ appearing (until now $\beta$ was just a dummy index)
in (\ref{a}) distinguishes a length scale, which is
called the thermal wave-length. 
In fact, we can turn the argument up side down: given a thermal field theory 
$\O \to {\cal R}_\beta (\O)$,
it is not necessary to provide an explicit expression for the effective Hamiltonian $H_\beta$.
It is already uniquely specified by the pair $({\cal R}_\beta, \Omega_\beta)$:
by Stone's theorem 
there exists a unique self-adjoint generator~$H_\beta$ such that 
$\Delta = \exp (- \beta H_\beta )$.
Modular theory implies that for $0 \le \beta < \infty$ the operator~$H_\beta$
is not semi-bounded but its spectrum is symmetric and
consists typically of the whole real line~\cite{A 72}\cite{tBW}.
%
%
%

\vskip 1cm

\subsection{Doubling the degrees of freedom}

\noindent
We now present the first step of our construction, which can be understood as a local purification.  
Consider some $\delta > 0$ and two space-time regions $\O$ and $\hat{\O}$ such that
$\O + te \subset \hat{\O}$ for~$|t| < \delta $. 
In a forthcoming paper \cite{Jae d} we will show that
the so-called split property for the net of von Neumann algebras $\O \to {\cal R}_\beta (\O)$
follows from the nuclearity condition (\ref{nuc}). 
It asserts that there exists a type I factor ${\cal N}$ such that
\beq {\cal R}_\beta (\O) \subset {\cal N} \subset {\cal R}_\beta (\hat{\O}) . \label{split}
\eeq
\vskip -.2cm

\Rem{If the KMS state is locally normal w.r.t.\ the vacuum representation, then the
split property for the vacuum representation automatically implies the split property
for the thermal representation.}
  
The following result is a consequence of the split inclusion (\ref{split}).

\begin{lemma}
{Let $\O$ be an open and bounded space-time region.
Then the von Neumann algebra
\[   {\cal M}_\beta (\O) := {\cal R}_\beta (\O) \vee j \bigl( {\cal R}_\beta (\O) \bigr)  \]
is naturally isomorphic to the tensor product of
${\cal R}_\beta (\O) $ and $ j \bigl( {\cal R}_\beta (\O) \bigr)$. 
I.e., there exists a unitary operator 
$V \colon \H_\beta \to \H_\beta \otimes \H_\beta$ such that
\beq V{\cal M}_\beta (\O)V^* = {\cal R}_\beta (\O) \otimes j \bigl( {\cal R}_\beta (\O) \bigr) . 
\label{x} \eeq
}
\end{lemma}

\vskip .3cm

\Pr{The split property (\ref{split}) implies that there 
exists a product vector $\Omega_p \in \H_\beta$, cyclic and separating for 
${\cal R}_\beta (\O) \vee {\cal R}_\beta (\hat {\O})' $,
such that
\[  (\Omega_p \, , \, A B   \Omega_p) =  (\Omega_\beta \, , \, A \Omega_\beta) 
(\Omega_\beta \, , \, B \Omega_\beta) \]
for all $A \in {\cal R}_\beta (\O)$ and $B \in {\cal R}_\beta (\hat {\O})'$ \cite{Jae d}.
The product vector $\Omega_p$ can be utilized to define a linear operator $V \colon \H_\beta 
\to \H_\beta  \otimes \H_\beta$ by linear extension of
\beq  V A B   \Omega_p 
=  A \Omega_\beta \otimes B \Omega_\beta , \label{b} \eeq
where $A \in {\cal R}_\beta (\O) $ and 
$B \in {\cal R}_\beta (\hat{\O})'$.
The operator $V$ is unitary. Inspecting (\ref{b}) we find 
\beq V {\cal R}_\beta (\O) V^* = {\cal R}_\beta (\O) \otimes \unit  
\quad \hbox{and} \quad
 V {\cal R}_\beta (\hat{\O})' V^* = \unit \otimes {\cal R}_\beta (\hat{\O})' . \label{y}  \eeq  
The inclusion $j \bigl( {\cal R}_\beta (\O) \bigr) \subset
{\cal R}_\beta (\hat{\O})'$ implies that the von Neumann algebra
${\cal M}_\beta (\O)$
is naturally isomorphic to the tensor product of
${\cal R}_\beta (\O) $ and $ j \bigl( {\cal R}_\beta (\O) \bigr)$ and the relation (\ref{x}) is 
a consequence of (\ref{y}).} 

\Rem{The algebras ${\cal R}_\beta (\O) $ 
and $ j \bigl( {\cal R}_\beta (\O) \bigr)$ are weakly statistically independent, i.e., 
$0 \ne A \in {\cal R}_\beta (\O) $ 
and $0 \ne B\in j \bigl( {\cal R}_\beta (\O) \bigr)$ implies 
$AB \ne 0$ (Schlieder property) \cite{Jae d}.
In this sense one can speak of a doubling of degrees of freedom.} 

The elements of ${\cal M}_\beta (\O)$ will in general not show analyticity properties with respect 
to~$\Omega_\beta$. Thus it seems that the essence of a thermal field theory gets lost, 
when we `double the degrees of freedom' and consider the net 
$\O \to {\cal M}_\beta (\O) $
instead of the net of observables~$\O \to {\cal R}_\beta (\O)$. 
But, due to the natural tensor product 
structure of ${\cal M}_\beta (\O)$, we can recover certain 
analyticity properties w.r.t.\ $\Omega_p$ and a new auxiliary one-parameter group of unitary 
operators:

\Def{A one-parameter group of unitary operators $s \mapsto
\Delta_p^{-is} \colon \H_\beta \to \H_\beta$, $ s \in \R$,  
and an anti-unitary operator $J_p \colon \H_\beta \to \H_\beta$ are given  by linear extension of
\beq   \Delta_p^{-is} AB \Omega_p  :=  V^*  \bigl(  
 \Delta^{-is} A \Omega_\beta \otimes \Delta^{is} B \Omega_\beta \bigr), \qquad s \in \R, \label{deltap}\eeq
and, respectively, 
\[   J_p  AB \Omega_p  :=  V^*  \bigl(  
 J A \Omega_\beta \otimes J B \Omega_\beta \bigr) , \]
where $A \in {\cal R}_\beta (\O)$ and $B \in {\cal R}_\beta (\hat{\O})'$. }

By Stone's theorem there exists a unique self-adjoint operator $H_p$
such that
\[  \Delta_p  = {\rm e}^{- \beta H_p}  \qquad \hbox{and} \qquad H_p \Omega_p = 0 . \]
The vector $\Omega_p \in \H_\beta$ is cyclic and separating for 
${\cal R}_\beta (\O) \vee {\cal R}_\beta (\hat {\O})' $. It follows from 
the definition (\ref{b}) of $V$ and the Reeh-Schlieder property 
of $\Omega_\beta$ that the product vector $\Omega_p$ is cyclic (and of course 
separating) for~${\cal M}_\beta (\O)$ too. 

\begin{theoreme}
{Let $\O_\circ$ and $\O$ be two space-time regions 
such that
$ \O_\circ + te \subset \O$  for~$|t| < \delta_\circ$. Then
\vskip .3cm
\halign{ \indent \indent \indent #  \hfil & \vtop { \parindent =0pt \hsize=34.8em
                            \strut # \strut} \cr 
{\rm (i)}    & $\Delta_p^{-is}$ respects the local 
structure of $\M_\beta (\O)$ for $|s|$ sufficiently small,  
i.e.,
\beq \Delta_p^{-is} \M_\beta (\O_\circ) \Delta_p^{is} \subset \M_\beta (\O_\circ + s \beta \cdot e) 
\qquad \forall |s \beta | < \delta_\circ . \label{split2} \eeq
\cr
{\rm (ii)} & 
the group of unitaries $s \mapsto \Delta_p^{-is}$ coincides for $a \in \A(\O_\circ)$ and $|s \beta | < \delta_\circ$---up to rescaling---with the time-evolution, i.e.,
\[ Ê\Delta_p^{-is} \pi_\beta (a) \Delta_p^{is} 
= \pi_\beta \bigl( \tau_{s \beta} (a) \bigr) \]
for $ |s \beta | < \delta_\circ $. 
\cr}
}
\end{theoreme}

\vskip .3cm

\Pr{The inclusion (\ref{split2}) follows from the definition (\ref{deltap})
of $\Delta_p$ and the inclusions 
\[   \hat \tau_{t } \bigl( {\cal R_\beta} (\O_\circ) \bigr) 
\subset {\cal R_\beta}  (\O_\circ + t e)
\quad \hbox {and} \quad 
\hat \tau_{t } \bigl( j ( {\cal R_\beta} (\O_\circ) ) \bigr) \subset j \bigl(
{\cal R_\beta}  (\O_\circ + t e) \bigr), \]
which hold for $|t | < \delta_\circ$.}

\begin{lemma}
{Consider some $\delta > 0$ and two space-time regions $\O$ and $\hat{\O}$ such that
$\O + te \subset \hat{\O}$ for~$|t| < \delta $. 
Let $\Omega_p $ and $ \Delta_p$ be the product vector specified in (\ref{b}) and the operator defined in~(\ref{deltap}).
Then
$ {\cal M}_\beta (\O) \Omega_p $ is in the domain~$   {\cal D} \bigl(\Delta_p^{ \alpha} \bigr)$ 
of $\Delta_p^{ \alpha}$ for $0 \le \alpha \le 1 / 2$. Moreover,
the identity $J_p \Delta_p^{1/2} M \Omega_p = M^* \Omega_p $
holds true for  all $M \in {\cal M}_\beta (\O)$.}
\end{lemma}

\vskip .3cm

\Pr{By definition,
$J_p^2 = \unit $, $J_p \Omega_p = \Omega_p$ and
$$ {  J_p \Delta_p^{1/2} AB \Omega_p 
 =  V^*  \bigl(  
A^* \Omega_\beta \otimes  B^* \Omega_\beta \bigr)
= A^* B^* \Omega_p
= (AB)^* \Omega_p }$$
for all  $A \in {\cal R}_\beta (\O)$ and 
$B \in j \bigl( {\cal R}_\beta (\O) \bigr)$. 
Since 
$ p^\alpha  \le 
\max (1 , p) < 1 + p $ for $0 \le \alpha \le 1$
and $p > 0$, the spectral resolution of the positive operator $\Delta_p^{1/2}$ 
implies that~${\cal M}_\beta (\O) \Omega_p \subset   {\cal D} \bigl(\Delta_p^{ \alpha} \bigr)
$ for~$0 \le \alpha \le 1 / 2$.
}

Nevertheless, $J_p$ and $\Delta_p$ are {\sl not}
the modular objects associated to $\bigl( {\cal M}_\beta (\O) , \Omega_p \bigr)$.

\begin{theoreme}
Let $\O_\circ$ and $\O$ be two space-time regions 
such that
$ \O_\circ + te \subset \O$  for~\hbox{$|t| < \delta_\circ$}. Then the inclusion of von Neumann algebras 
${\cal M}_\beta (\O_\circ) \subset {\cal M}_\beta (\O)$ is  a standard split
inclusion and there exists a unitary operator $W \colon \H_\beta \to \H_\beta \otimes \H_\beta$ 
such that
\[  W {\cal M}_\beta (\O_\circ) W^* = {\cal M}_\beta (\O_\circ) \otimes \unit 
\quad \hbox{and} \quad
 W {\cal M}_\beta (\O )' W^* = \unit \otimes {\cal M}_\beta (\O)' . \]   
\end{theoreme}
\noindent(A split in\-clusion $ {\cal A} \subset {\cal B}$ is called standard  (see \cite{DL}), if there exists
a vector $ \Omega$ which is cyclic for~$ {\cal A}' \wedge {\cal B}$ 
as well as for~$ {\cal A}$ and~$ {\cal B}$.)

\vskip .3cm

\Pr{From the split inclusions
\[  {\cal R}_\beta (\O_\circ) \subset {\cal N}_\circ \subset {\cal R}_\beta (\O ) 
\quad \hbox{and} \quad
j \bigl( {\cal R}_\beta (\O_\circ) \bigr) \subset 
j \bigl( {\cal N}_\circ \bigr) \subset 
j \bigl( {\cal R}_\beta (\O ) \bigr)   \]
we infer that there exists a type I factor, namely ${\cal N}_\circ \vee 
j \bigl( {\cal N}_\circ \bigr)$, 
such that
\[  {\cal M}_\beta (\O_\circ) \subset {\cal N}_\circ \vee j \bigl( {\cal N}_\circ \bigr)
\subset {\cal M}_\beta (\O)  . \]
All infinite type I factors with infinite commutant on 
the separable Hilbert space
$\H_\beta$ are unitarily
equivalent to $\B(\H_\beta) \otimes \unit  \, $ (\cite{KR}, Chapter 9.3). 
Thus there exists a unitary operator $W \colon \H_\beta \to \H_\beta \otimes \H_\beta$ such that 
$ {\cal N}_\circ \vee j \bigl( {\cal N}_\circ \bigr) 
= W^* \bigl( \B(\H_\beta)  \otimes \unit  \, \bigr) W $.
Now consider $\omega_\beta ( \, . \, ) := (\Omega_\beta \, , \, .\, \Omega_\beta)$ 
and $\omega_p ( \, . \, ) := (\Omega_p \, , \, .\, \Omega_p)$
as two normal states over 
${\cal M}_\beta (\O_\circ)$ and~${\cal M}_\beta (\O)'$, 
respectively. Set
\[  \phi_p (C) := (\omega_\beta \otimes \omega_p) (W C W^*) \qquad \forall
C \in {\cal M}_\beta (\O_\circ) \vee {\cal M}_\beta (\O)'. \]
Then $\phi_p$ is a normal state over ${\cal M}_\beta (\O_\circ) \vee {\cal M}_\beta (\O)'$,
which satisfies 
$ \phi_p (M N) = \omega_\beta (M) \cdot \omega_p (N)  $
for all $M \in {\cal M}_\beta (\O_\circ)$ and $N \in {\cal M}_\beta (\O)'$.
In the presence of a separating vector each normal 
state is a vector state (\cite[7.2.3]{KR}). In fact, there exists a unique 
vector $\eta$ in the natural positive 
cone ${\cal P}^\natural \bigl({\cal M}_\beta (\O_\circ) \vee {\cal M}_\beta (\O)', \Omega_\beta
\bigr) $
such that 
\[  (\eta \, , \, MN \eta) = \phi_p (M N) =   (\Omega_\beta \, , \, M \Omega_\beta) 
(\Omega_p \, , \, N \Omega_p)  \]
for all $M \in {\cal M}_\beta (\O_\circ)$ and $N \in {\cal M}_\beta (\O)'$ 
(\cite[2.5.31]{BR}).  Thus the operator $W \colon \H_\beta \to 
\H_\beta \otimes \H_\beta$ can now be specified by linear extension of
\beq  W M N \eta  =  M \Omega_\beta \otimes N \Omega_p   , \label{pv} \eeq
where $M \in {\cal M}_\beta (\O_\circ)$ and $N \in {\cal M}_\beta (\O)'$.
Consequently,
\[  W {\cal M}_\beta (\O_\circ) W^* = {\cal M}_\beta (\O_\circ) \otimes \unit 
\quad \hbox{and} \quad
 W {\cal M}_\beta (\O )' W^* = \unit \otimes {\cal M}_\beta (\O)' . \]  
The vector $\Omega_\beta$ is cyclic and separating for  ${\cal M}_\beta (\O_\circ)$ and the vector
$\Omega_p$ is cyclic and separating for~${\cal M}_\beta ( \O  )'$. Thus the vector~$\Omega_\beta \otimes
\Omega_p$ is cyclic and separating for  ${\cal M}_\beta (\O_\circ)
\otimes {\cal M}_\beta ( \O  )' $ and the split inclusion ${\cal M}_\beta (\O_\circ) \subset {\cal M}_\beta (\O)$ is standard.}

\vskip 1cm

\section{Localized excitations of a KMS state}

Taking the auxiliary structure developed in the previous section into account, 
we can now adapt the method of Buchholz and Junglas 
to  thermal representations. 

\subsection{Consequences of the nuclearity condition}

Imposing strict localization on an excitation (see Proposition 3.3 (iii) below)
does not lead to a convenient notion. The split property provides the key 
to a more convenient definition of a localized excitation. 
However, it leaves a lot of freedom, for instance one could request additional properties 
for some subregion in $\O_\circ \cap \hat \O$. In this sense the follwing definition only provides one possible choice,
fixed by choosing a specific product vector $\eta$.

\noindent

\Def{Let $\O_\circ , \O$ and $ \hat {\O}$ denote three space-time regions 
such that for some $\delta_\circ$, $\delta >0$
\beq \O_\circ + te \subset \O  \quad \forall |t| < \delta_\circ
\quad
\hbox{and}
\quad
\O + te \subset \hat {\O}  \quad \forall |t| < \delta . \label{regions}
\eeq
The Hilbert space $\H_\Lambda  \subset \H_\beta$,
$\Lambda := (\O_\circ , \O , \hat {\O})$, of localized excitations of the 
KMS state~$\omega_\beta$ is given by 
\beq \H_\Lambda  
:= \overline {  {\cal M}_\beta ( \O_\circ )   \eta  }. \label{a1} \eeq
The projection onto $\H_\Lambda$ is denoted by $E_\Lambda$.}

\noindent{\sl Notation.} Here
${\cal M}_\beta ( \O_\circ )$ denotes the von Neumann algebra generated by
${\cal R}_\beta ( \O_\circ )$ and $j \bigl( {\cal R}_\beta ( \O_\circ ) \bigr)$ and
$\eta \in \H_\beta$ denotes the unique\footnote
{\rm Fixing the product vector with respect to some natural positive 
cone is mathematically convenient,
but not necessary. In fact, we expect that  
different `boundary conditions' are realized by different choices of $ \eta$.
In the thermodynamic limit different choices of the boundary conditions  might lead to different phases.}
product vector in the natural positive 
cone ${\cal P}^\natural \bigl({\cal M}_\beta (\O_\circ) \vee {\cal M}_\beta (\O)', \Omega_\beta
\bigr) $
satisfying
\beq  (\eta \, , \, MN \eta) =   (\Omega_\beta \, , \, M \Omega_\beta) 
(\Omega_p \, , \, N \Omega_p)  \label{a2} \eeq
for all $M \in {\cal M}_\beta (\O_\circ)$ and $N \in {\cal M}_\beta (\O)'$. 
As before, $\Omega_p$ denotes the unique product vector in the natural positive 
cone ${\cal P}^\natural \bigl({\cal R}_\beta (\O) \vee {\cal R}_\beta (\hat {\O})', \Omega_\beta
\bigr) $
satisfying
\beq  (\Omega_p \, , \, A B   \Omega_p) =  (\Omega_\beta \, , \, A \Omega_\beta) 
(\Omega_\beta \, , \, B \Omega_\beta) \label{a3} \eeq
for all $A \in {\cal R}_\beta (\O)$ and $B \in {\cal R}_\beta (\hat {\O})'$.
\vskip .5cm
 
Note that $W$--- as specified in (\ref{pv})---is unitary and 
$W M N W^*  =  M
\otimes N $
for $M \in {\cal M}_\beta (\O_\circ)$ and $N \in {\cal M}_\beta (\O)'$.
Using the isometry $W$ we can write
\[ \H_\Lambda
 =  W^*  \overline { {\cal M}_\beta (\O_\circ)  \Omega_{\beta} 
\otimes  \Omega_p}
=  W^* ( \H_{\beta} 
\otimes  \Omega_p )  \]
and $E_\Lambda  =
W^*  ( \unit \otimes P_{\Omega_p}) W $. Here $P_{\Omega_p} \in \B (\H_\beta)$ denotes the 
projection onto $ \C \cdot \Omega_p$.

\vskip .5cm
The following proposition summarizes the properties of the Hilbert space $\H_\Lambda$. It justifies 
the claim stated at the beginning of this subsection.

\goodbreak
\begin{proposition}
{Given a triple $\Lambda := ( \O_\circ, \O, \hat {\O})$ of space-time regions
as specified in {\rm (\ref{regions})} we find:
\vskip .3cm
\halign{ \indent \indent \indent #  \hfil & \vtop { \parindent = 0pt \hsize=34.8em
                            \strut # \strut} 
\cr 
(i)    & The Hilbert space $\H_\Lambda$ is invariant under the action of elements of
       ${\cal M}_\beta (\O_\circ)$, i.e.,
$ {\cal M}_\beta (\O_\circ) \H_\Lambda  = \H_\Lambda $.
\cr
(ii)    & Vectors  in $ \H_\Lambda$ induce product states for the pair 
$\bigl({\cal M}_\beta (\O_\circ) , \M_\beta (\O)' \bigr)$: if $\Psi \in \H_\Lambda$, then  
\[ (\Psi \, , \, MN \Psi) = (\Psi\, , \, M \Psi) (\Omega_p \, , \, N \Omega_p) \]
for all  
$M \in {\cal M}_\beta (\O_\circ)$ and $N \in \M_\beta (\O)'$.
\cr
(iii)    & The vector states associated with $ \H_\Lambda$ represent strictly localized
excitations of the KMS state, i.e., they coincide with the original KMS state $\omega_\beta$
in the space-like complement of $\hat {\O}$:  if $\Psi \in \H_\Lambda$, then  
\[ (\Psi \, , \, \pi_\beta (a) \Psi) = \omega_\beta (a) 
\qquad \forall a \in \A^c (\hat{\O}) . \]
Here $\A^c (\hat{\O})  $ denotes the $C^*$-algebra  generated by 
$\{ a \in \A : [a , b ] = 0 \, \, \forall b \in \A(\hat{\O}) \}$
and {\rm not} the commutant of $\pi_\beta \bigl( \A(\hat{\O})  \bigr)$ in $\B(\H_\beta)$. 
\cr
(iv)    & $\H_\Lambda$ is complete in the following sense: to every normal state $\phi$
on ${\cal M}_\beta (\O_\circ)$ there exists a $\Phi \in \H_\Lambda$ such that
$(\Phi\, , \, M \Phi) = \phi(M) $
for all $M \in {\cal M}_\beta (\O_\circ)$.
\cr} }
\end{proposition}

\vskip .3cm

\Pr {We simply adapt the proof of the corresponding result 
by Buchholz and Junglas to our situation:
\vskip .2cm
\noindent
\halign{ \indent \indent \indent # \hfil & \vtop { \parindent =0pt \hsize=34.8em
                            \strut # \strut} \cr 
(i)  & follows from the definition;
\cr
(ii) &
follows from (\ref{a1}) and (\ref{a2});
\cr
(iii) &
follows from (\ref{a1}), (\ref{a2}) and (\ref{a3}).
\cr
(iv) 
& 
Since ${\cal M}_\beta (\O_\circ)$ has a cyclic and separating vector, there exists a vector 
$ \tilde {\Phi} \in \H_\beta$ which induces the given normal state
$\phi$ on  ${\cal M}_\beta (\O_\circ) $.
It follows that  the vector $\Phi := W^* ( \tilde {\Phi} 
\otimes  \Omega_p )  \in \H_\Lambda $ satisfies (iv).
\cr}}

We need one more lemma, 
in order to show that the restriction of the operator $\Delta_p^{\alpha}$
to the subspace $\H_\Lambda$ is  trace class
for $ 0 < \alpha < 1  / 2$.

\begin{lemma}
{Assume that the nuclearity condition (\ref{nuc}) holds true.
It follows that 
\vskip .2cm
\noindent
\halign{ \indent \indent \indent # \hfil & \vtop { \parindent =0pt \hsize=34.8em
                            \strut # \strut} \cr 
(i)  & the maps  $\vartheta_{\alpha, {\cal O}} \colon {\cal M}_\beta (\O)  \to \H_\beta$,
\[  M  \mapsto 
\Delta_p^{\alpha} M  \Omega_p ,  \qquad 0 \le \alpha \le 1/2, \]
are nuclear for $0 < \alpha < 1/2$; 
%
%
\cr
(ii) & the nuclear norm of $\vartheta_{\alpha, {\cal O}}$ is bounded by  
\[  \|  \vartheta_{\alpha, {\cal O}}  \|   \le 
{\rm e}^{2 c r^d \bigl 
( \alpha^{-m} + (1 / 2 - \alpha)^{-m} \bigr) } ,
\quad \quad c, m, d > 0, \]
where $r$ denotes the diameter of $\O$ and $c, m, d$ 
are the constants appearing in the bound~(\ref{nuc}) on the nuclear norm of the map $\Theta_{\alpha, \O}$.
\cr}}
\end{lemma}

\vskip .3cm

\Pr{Let $A \in {\cal R}_\beta (\O)$ and $B \in j \bigl( {\cal R}_\beta (\O) \bigr)$. 
By definition,
\[  \vartheta_{\alpha, {\cal O}} (AB)   
=  V^*  \bigl( \Delta^\alpha A \Omega_\beta  \otimes \Delta^{-\alpha} B \Omega_\beta \bigr) . \]
The maps
$ A \mapsto \Delta^\alpha  A  \Omega_\beta$
and
$B \mapsto \Delta^{ - \alpha } B \Omega_\beta$
are nuclear for $0 < \alpha < 1 /2$. The tensor product of two nuclear maps itself  is  a
nuclear map and the norm is bounded by the product of the nuclear norms \cite{P}. }

\begin{proposition}
{Let $\Lambda( \O_\circ , \O, \hat {\O} )$ be a triple of space-time regions
as specified in (\ref{regions}). Assume the nuclearity condition (\ref{nuc}) holds true.
It follows that 
the operator $\Delta_p^{\alpha} E_\Lambda $,
acting on the Hilbert space $\H_\beta$, is of trace-class for $0 < \alpha < 1  / 2$, and 
\[ \Tr  \, | \Delta_p^{\alpha} E_\Lambda |  
\le
{\rm e}^{2 c r^d \bigl 
( \alpha^{-m} + (1 / 2 - \alpha)^{-m} \bigr) } ,
\qquad  c, m, d > 0, \]
where $r$ denotes the diameter of $\O$ and $c, m, d$ 
are the constants appearing in the bound
(\ref{nuc}) on the nuclear norm of the map $\Theta_{\alpha, \O}$.}
\end{proposition}
 
\vskip .3cm
 
\Pr{The proof of this proposition is more or less identical to the one given by
Buchholz and Junglas \cite{BJu 89} for the vacuum case. We present it for completeness only. 
\vskip .1cm
\noindent
i.) The first step is to construct a convenient orthonormal basis of $\H_\Lambda$. 
Let $\{ \Psi_i \}_{i \in \n} $ be an orthonormal basis of $\H_\beta$ with
$\Psi_1 = \Omega_\beta$. Set
\beq U_{i,j} : = W^*  (M_{i,j} \otimes \unit ) W  , \label{b1} \eeq
where $M_{i,j} \in \B ( \H_\beta )$ are matrix units given by
\[  M_{i,j} \Psi : = (\Psi_j, \Psi) \Psi_i
\qquad \forall \Psi \in \H_\beta. \]
Since $W^*  \bigl( \B ( \H_\beta ) \otimes \unit  \bigr) W =
{\cal N}_\circ \vee j \bigl( {\cal N}_\circ \bigr)$, we infer from (\ref{b1}) that 
$U_{i,j} \in {\cal N}_\circ \vee j \bigl( {\cal N}_\circ \bigr) $.
Furthermore, 
\[  U_{i,j}^* = U_{j,i}  , 
\quad
U_{i,j}  U_{k,l} = \delta_{j,k} U_{i,l}  , 
\quad
\hbox{and}
\quad
\slim_{ N \to \infty } \sum_{i = 1 }^N  U_{i,i} = \unit \, . \]
Combining (\ref{pv}) and (\ref{b1}) we find
$U_{i,1} \eta = W^*  (\Psi_i \otimes \Omega_p) $.
Thus $\{   U_{i,1} \eta \}_{i \in \n}$ is the desired orthonormal basis of $\H_\Lambda $.
Note that (\ref{pv}) holds true for all $N \in \bigl( {\cal N}_\circ 
\vee j ( {\cal N}_\circ) \bigr)' $. Therefore $\| N \eta \| = \| N \Omega_p \|$. Consequently, 
we can introduce an isometry $I \in {\cal N}_\circ \vee j \bigl( {\cal N}_\circ \bigr)$ by setting
\[   N \eta = I N \Omega_p \qquad  \forall N \in \bigl( {\cal N}_\circ 
\vee j ( {\cal N}_\circ) \bigr)' . \] 
We can now represent the orthonormal basis  $\{   U_{i,1} \eta \}_{i \in \n}$   by vectors
$\Gamma_i := U_{i,1} \eta  = U_{i,1} I \Omega_p $, 
where $U_{i,1} I \in {\cal N}_\circ \vee j \bigl( {\cal N}_\circ \bigr)
\subset {\cal M}_\beta (\O)$. It follows that
$\Gamma_i \in {\cal D} \bigl( \Delta_p^{\alpha} \bigr)$ for $0 < \alpha < 1 /2$ and~$i \in \N$.
Especially, $\eta =: \Gamma_1 \in {\cal D} \bigl( \Delta_p^{\alpha} \bigr)$ 
for $0 < \alpha < 1 /2$.
\vskip .2cm 
\noindent
ii.) Polar decomposition of the closeable operator $\Delta_p^{\alpha} E_\Lambda$ yields
$\Delta_p^{\alpha} E_\Lambda
= F \cdot | \Delta_p^{\alpha} E_\Lambda |$, 
where~$F$ is a partial isometry with range in $\H_\Lambda$.
Introducing a set of linear functionals $\phi_i$ (which can be chosen to be continuous with
respect to the ultra-weakly topology induced by~${\cal M}_\beta (\O)$ \cite{BD'AL 90b})
and vectors $\Phi_i \in \H_\beta$ corresponding to the 
nuclear map  $\vartheta_{\alpha, {\cal O} }$
we obtain
\begin{align*}
\Tr \, | \Delta_p^{\alpha} E_\Lambda |
&= \sum_i ( U_{i,1} I  \Omega_p \, , \, 
F^*  \Delta_p^{\alpha}     
U_{i,1} I \Omega_p ) \\
&= \sum_i \bigl( U_{i,1} I  \Omega_p \, , \,  F^*  
\vartheta_{\alpha, {\cal O} } ( U_{i,1} I ) \bigr) \\
&= \sum_i \sum_n     \phi_n ( U_{i,1} I ) 
\cdot  ( U_{i,1} I \Omega_p \, , \,  F^* \Phi_n)  \\
&\le \sum_i \sum_n   | \phi_n ( U_{i,1} I ) |
\cdot   \|    U_{1,i}  F^*  \Phi_n  \| . 
\end{align*}
Buchholz and Junglas  have shown the following inequality \cite{BJu 89}:
\[ 
\sum _i  | \psi ( U_{i,1} ) | \cdot
\| U_{1,i} \Psi \|
\le  \| \psi \| \,  \|  \Psi \|, 
\]
for $\Psi \in \H_\beta$ and~$\psi$ an ultra-weakly continuous linear
functional on~${\cal M}_\beta (\O)$. Consequently,
$\Tr \, | \Delta_p^{\alpha} E_\Lambda | \le  \sum_n   \| \phi_n \| \,  \| \Phi_n  \| $. 
Taking the infimum with respect to all decompositions of the respective nuclear maps
we find
$\Tr \,  | \Delta_p^{\alpha} E_\Lambda |
\le   \| \vartheta_{\alpha, {\cal O} } \| $. 
}

\vskip .5cm

\subsection{Local KMS states for a new temperature}

\noindent
Proposition 3.3 allow us
to define ``local quasi-Gibbs'' states, which are {\sl local} $(\tau, \beta')$-KMS states
for the new temperature $1/ \beta'$ in
the interior of $\O_\circ$ and $(\tau, \beta)$-KMS states
for the original temperature $1/\beta$ outside of 
$\hat {\O}$. Before we do so, 
we give a precise meaning to the statement
that a local excitation $\omega_\Lambda$ of a KMS state $\omega_\beta$
satisfies a {\sl local KMS condition} for the new  
temperature~$1/ \beta'$ in a bounded region 
$\O_\circ$. Note that any $\beta'$ ($0 < \beta' < \infty$) can be decomposed
into some $\alpha$ ($0 < \alpha < 1/2$) and some (minimal) $n \in \N$ such that $\beta' = \alpha n \beta$.

\Def{Let $\beta' > 0$ and let $n \in \N$ be the smallest natural number such that 
$n \alpha \beta = \beta' $ for some $\alpha$ $(0 \le \alpha \le 1 / 2)$.  
A state $\omega_\Lambda $
satisfies the {\sl local KMS condition} at temperature $1/\beta'$
in some bounded space-time region~$\O_\circ \subset \R^4$ if for any 
subregion $\O_{\circ \circ} \subset \O_\circ$
whose closure is contained in the interior of~$\O_\circ$ there exists some $\delta_{\circ \circ} >0$ 
and a 
function $F_{a,b}$ for every pair 
of operators $a$, $b \in \A (\O_{\circ \circ})$ such that
\vskip .2cm
\halign{ \indent \indent \indent #  \hfil & \vtop { \parindent =0pt \hsize=35.8em
                            \strut # \strut} \cr 
(i)  & $F_{a,b}$ is defined on 
\[ {\cal G}_{n , \alpha} := \{z \in \C \mid 0 < \Im z < n \alpha \beta \}
\setminus  \{ z \in \C \mid | \Re z | \ge \delta_{\circ \circ}, \Im z =  k \alpha \beta, \, 
k = 1, \ldots, n-1 \}; \]
\vskip -.2cm
\cr
(ii)    & $F_{a,b}$ is bounded and analytic in its domain of definition; 
\cr
(iii)   &  $F_{a,b}$ is continuous for 
           $ \Im z  \searrow k \alpha \beta$ and $\Im z  \nearrow k \alpha \beta$, 
$k = 1, \ldots, n-1$; 
\cr
(iv)   &  $F_{a,b}$ is continuous at the boundary for 
$ \Im z  \searrow 0 $ and $\Im z  \nearrow n \alpha \beta$; 
\cr
(v)    & The respective boundary values are 
\[  
\label{almost} F_{a,b} (t) = \omega_\Lambda \bigl( a \tau_t(b) \bigr)
\hbox{ and }  F_{a,b} (t + i n \alpha \beta) = \omega_\Lambda \bigl( \tau_t(b) a \bigr)  
\hbox{  for }  |t| < \delta_{\circ \circ}. \]
\cr}
}

\vskip -.8cm

\Rem{To heat up the system locally is quite simple: For $\beta' < \beta / 2$ we find $n= 1$, i.e.,
no cuts appear in
${\cal G}_{1 , \alpha} = \{z \in \C \mid 0 < \Im z <  \alpha \beta \}$. To cool 
down the system locally is more delicate. One needs at least $n$ cuts, where
$n$ is the minimal natural number such that~$\beta' = n \alpha \beta$ $(0 < \alpha < 1/2)$.
Whether or not it is useful to operate  with more cuts then necessary is unknown to us.}

\begin{proposition}
{Let $\Lambda:= ( \O_\circ , \O, \hat {\O} )$ be a triple of space-time regions,
as specified in~(\ref{regions}). Let $n\in \N$ be the minimal natural
number such that $\beta' = n \alpha \beta$, $0 < \alpha < 1/2$.
Set, for $ n $ and~$ \alpha $ fixed,
\beq  \label{lKMS}
\rho_\Lambda := \frac{  \bigl( E_\Lambda \Delta_p^{\alpha} E_\Lambda \bigr)^n   
}{ \Tr \, \bigl( \Delta_p^{\alpha} E_\Lambda \bigr)^n }
\quad \hbox {and} \quad \omega_\Lambda  (a) 
:=  \Tr \, \rho_\Lambda \pi_\beta (a)  
\quad  \forall a \in \A . \eeq
Then $\rho_\Lambda$ is a density matrix, i.e., $\rho_\Lambda> 0$ and $\Tr \, \rho_\Lambda =1$,
and the following statements hold true:
\vskip .3cm
\halign{ \indent \indent \indent #  \hfil & \vtop { \parindent =0pt \hsize=34.8em
                            \strut # \strut} \cr 
(i)  & The states $\omega_\Lambda$ are product states, which coincide with
the given KMS state $\omega_\beta$ in the space-like complement of $\hat{\O}$; i.e.,
\[  \omega_\Lambda  ( a b') 
= \omega_\Lambda  (a)  \, 
\omega_\beta (b') \]
for all $a \in \A(\O_\circ)$ and $b' \in \A^c (\hat {\O})$. As before,
$\A^c (\hat{\O})  $ denotes the $C^*$-algebra  generated by 
$\{ a \in \A \mid [a , b ] = 0 \, \, \forall b \in \A(\hat{\O}) \}$. 
\cr
(ii)   &  The states $\omega_\Lambda$ are local $(\tau, n \alpha \beta)$-KMS states 
for the space-time region $\O_\circ$.  
\cr}  
}
\end{proposition}

\vskip .3cm

\Rem{For $\O_\circ ,  \O, \hat{\O} \to \R^4$ the denominator in (\ref{lKMS}) might go to $\infty$ or $0$. In any case we will 
leave the representation: we will have no operator convergence, neither in the weak nor in the strong sense and therefore 
we can only 
rely on  expectation values. After performing the thermodynamic limit, we will use these expectation values 
to  construct a 
new representation and check whether 
the new state satisfies the KMS condition \cite{Na}. We will see that it
will do so, under the assumptions we have imposed on the phase-space properties of our thermal field theory.}

\Pr{(i) Let $a \in \A(\O_\circ)$ and $b' \in \A^c (\hat {\O})$ and let
$P_{\Omega_p}$ denote the projection onto~$\C \cdot \Omega_p$. Since
$E_\Lambda \in {\cal M}_\beta (\O_\circ)' \subset \pi_\beta \bigl( \A( \O_\circ ) \bigr)'$, 
it follows that $[ E_\Lambda , \pi_\beta (a) ] = 0$.
Moreover, $E_\Lambda = W^* (\unit \otimes P_{\Omega_p}) W$
implies
\[  E_\Lambda \pi_\beta(b') E_\Lambda =
\omega_\beta ( b') E_\Lambda \qquad \forall b' \in \A^c (\hat{\O}) . \]
Using the cyclicity of the trace  
we find
\begin{align*}
\omega_\Lambda  ( ab')
&= \frac{ \Tr \bigl( E_\Lambda \Delta_p^{\alpha} E_\Lambda \bigr)^n \, 
\pi_\beta (a)  E_\Lambda \pi_\beta (b') E_\Lambda
}{ \Tr \bigl( \Delta_p^{\alpha} E_\Lambda \bigr)^n }
\\
&=\omega_\Lambda  ( a )  \omega_\beta(b') . 
\end{align*} 
\vskip .1cm
\noindent
(ii) Consider the case $n = 2$. Let $\delta_{\circ \circ} > 0$ and 
$\O_{\circ \circ}$ be an open space-time region
such that $\O_{\circ \circ} + t e \subset \O_{\circ}$ for~$|t| < \delta_{\circ \circ}$.
Let $a, b \in \A(\O_{\circ \circ})$. By assumption,   
$a \tau_t (b) \in \A (\O_\circ)$ for  $ |t | < \delta_{\circ \circ} $. 
Set
\[ F_{a,b}^{(1)} (z ) := 
\frac{ \Tr \, \pi_\beta (a) E_\Lambda    
\Delta_p^{-iz / \beta} \pi_\beta ( b )
\Delta_p^{\alpha + i z / \beta}   E_\Lambda 
\Delta_p^{\alpha} E_\Lambda 
}{
\Tr \, \bigl( \Delta_p^{\alpha} E_\Lambda \bigr)^2 }   \]
for $0 < \Im z  < \alpha \beta$. The function $F_{a,b}^{(1)} (z)$ is analytic in its domain and 
continuous at the boundary. We recall that 
$\Delta_p^{-it / \beta } \pi_\beta (b) \Delta_p^{it / \beta} = \pi_\beta \bigl( \tau_{t} (b) \bigr)
\in \pi_\beta \bigl( \A(\O_\circ) \bigr)$ for $|t | < \delta_{\circ \circ} $. 
Using once again 
the cyclicity of the trace and $E_\Lambda \in \pi_\beta \bigl( \A(\O_\circ) \bigr)'$, 
we conclude that 
\begin{align*}
\lim_{\Im z \searrow 0} F_{a,b}^{(1)} (z)  & = \frac{ \Tr \, \pi_\beta (a) E_\Lambda    
\pi_\beta \bigl( \tau_{\Re z} (b) \bigr)
\Delta_p^{\alpha}   E_\Lambda 
\Delta_p^{\alpha} E_\Lambda 
}{
\Tr \, \bigl( \Delta_p^{\alpha} E_\Lambda \bigr)^2 }   
\\
& = \frac{ \Tr \,     
\pi_\beta \bigl( a \tau_{\Re z} (b) \bigr) \bigl( E_\Lambda
\Delta_p^{\alpha}   E_\Lambda \bigr)^2
}{
\Tr \, \bigl( \Delta_p^{\alpha} E_\Lambda \bigr)^2 }   
\qquad  \forall |{\Re z}| < \delta_{\circ \circ}.   
\end{align*}
Thus
\beq
\lim_{\Im z \searrow 0} F_{a,b}^{(1)} (z)  = \omega_\Lambda \bigl( a \tau_{\Re z} (b) \bigr)  \qquad 
\forall |{\Re z}| < \delta_{\circ \circ}. \label{c1} 
\eeq
On the other hand, for $|{\Re z}| < \delta_{\circ \circ}$,
\begin{align*}
\lim _{\Im z \nearrow \alpha \beta} F_{a,b}^{(1)} (z ) 
& = \frac { \Tr \, \pi_\beta (a) E_\Lambda    
\Delta_p^{\alpha} \pi_\beta \bigl( \tau_{\Re z} (b) \bigr)
 E_\Lambda 
\Delta_p^{\alpha} E_\Lambda
}{
\Tr \, \bigl( \Delta_p^{\alpha} E_\Lambda \bigr)^2 } 
\\
& = \frac{ \Tr \, \pi_\beta (a) E_\Lambda    
\Delta_p^{\alpha}  E_\Lambda \pi_\beta \bigl( \tau_{\Re z} (b) \bigr) 
\Delta_p^{\alpha} E_\Lambda 
}{
\Tr \, \bigl( \Delta_p^{\alpha} E_\Lambda \bigr)^2 }. 
\end{align*}
For $ \alpha \beta < \Im z <  2 \alpha \beta$ we set 
\[ F_{a,b}^{(2)} (z ) := \frac{ \Tr \, \pi_\beta (a) E_\Lambda    
\Delta_p^{\alpha} E_\Lambda  
\Delta_p^{- \alpha -iz / \beta} \pi_\beta (b)  
\Delta_p^{2\alpha + i z / \beta}  E_\Lambda  
}{
\Tr \, \bigl( \Delta_p^{\alpha}  E_\Lambda \bigr)^2 } . 
\]
The function $F_{a,b}^{(2)} (z )$ is analytic in its domain and 
continuous at the boundary. 
By definition,
\[ \lim_{\Im z \searrow \alpha \beta} F_{a,b}^{(2)} (z ) 
= 
\frac{ \Tr \, \pi_\beta (a) E_\Lambda    
\Delta_p^{\alpha} E_\Lambda  
\pi_\beta \bigl( \tau_{\Re z} (b) \bigr)
\Delta_p^{\alpha}   E_\Lambda  
}{
\Tr \, \bigl( \Delta_p^{\alpha} E_\Lambda \bigr)^2 } 
= \lim_{\Im z \nearrow \alpha \beta} F_{a,b}^{(1)} (z) 
\quad \forall |{\Re z}| < \delta_{\circ \circ}. \]
Furthermore,  $F_{a,b}^{(2)}$ satisfies  
\begin{align*}
\lim_{ \Im z \nearrow 2 \alpha \beta} F_{a,b}^{(2)} (z) 
& = \frac{ \Tr \, \pi_\beta (a) E_\Lambda    
\Delta_p^{\alpha} E_\Lambda
\Delta_p^{\alpha}  \pi_\beta \bigl( \tau_{\Re z} (b) \bigr)  
E_\Lambda  
}{
\Tr \, \bigl( \Delta_p^{\alpha} E_\Lambda \bigr)^2 }
\\
& = \frac{ \Tr \, \pi_\beta (a) \bigl( E_\Lambda    
\Delta_p^{\alpha} E_\Lambda \bigr)^2 \pi_\beta \bigl( \tau_{\Re z} (b) \bigr)  
}{
\Tr \, \bigl( \Delta_p^{\alpha} E_\Lambda \bigr)^2 }
\qquad \forall |{\Re z}| < \delta_{\circ \circ}. 
\end{align*}
Thus
\beq
\label{c2} 
\lim_{ \Im z \nearrow 2 \alpha \beta} F_{a,b}^{(2)} (z) 
= \omega_\Lambda  \bigl( \tau_{\Re z} (b) a  \bigr) 
\qquad \forall |{\Re z}| < \delta_{\circ \circ}. 
\eeq
Using the Edge-of-the-Wedge theorem \cite{SW} we conclude
that $F_{a,b}^{(1)}$ and $F_{a,b}^{(2)}$ are  
the restrictions to the upper (resp.\ lower)
half of the double cut  strip 
\[  {\cal G}_{2,\alpha} = \{ z \in \C \mid 0 < \Im z < 2 \alpha \beta \} \setminus 
\{ z \in \C \mid | \Re z | \ge \delta_{\circ \circ} , \Im z = \alpha \beta \}  \]
of a function 
\[ 
F_{a,b} (z) 
:= 
\left\{
\eqalign{
&  { F_{a,b}^{(2)} (z)  }
\cr
&
{ F_{a,b}^{(1)} (z) } 
}
\right\} 
\hbox{\rm for}
\left\{
\eqalign{
& { \alpha \beta < \Im z < 2 \alpha \beta ,}  
\cr
& {0 < \Im z < \alpha \beta,}  }  
\right\} 
\]
defined and continuous on the closure of ${\cal G}_{2, \alpha}$
and analytic for $z \in {\cal G}_{2, \alpha}$.
From (\ref{c1}) and~(\ref{c2}) we infer
$F_{a,b} (t) = \omega_\Lambda  \bigl( a \tau_{t} (b) \bigr)  
$ and $F_{a,b} (t + i 2 \alpha \beta ) 
= \omega_\Lambda  \bigl( \tau_{t} (b) a  \bigr) 
$ for $|t| < \delta_{\circ \circ}$. 
Analogous results for arbitrary $n \in \N$ can be established by the same line of arguments
but with considerable more effort.}

\vskip 1cm

\section{The thermodynamic limit}

We will now control the surface energies in the limit $\O_\circ, \O, \hat{\O} \to \R^4$. 
Since we do not have explicit expressions for the surface energies, our approach is quite involved.
The first step is to control the convergence of product vectors.

\subsection{Consequences of the cluster condition}

\noindent
Let us  introduce some
notation: Let $\Lambda_i = \bigl( \O_\circ^{(i)} , \O^{(i)} , \hat{\O}^{(i)} \bigr)$ be a sequence
of 
triples of double cones with diameters~$\bigl( r^{(i)}_\circ , r^{(i)} , \hat{r}^{(i)} \bigr)$. 
We consider the  product vectors~$\Omega_p^{(i)} $ and~$ \eta_i , \chi_i \in {\cal P}^\natural
\bigl({\cal M}_\beta ( \O_\circ^{(i)}) \vee {\cal M}_\beta ( \O^{(i)})' , 
\Omega_\beta \bigr)$,
which satisfy
\[  \bigl( \Omega_p^{(i)} \, , \, A B  \, \Omega_p^{(i)} \bigr) = (\Omega_\beta \, , \, A \Omega_\beta)
(\Omega_\beta \, , \, B \Omega_\beta)  \]
for $A \in {\cal R}_\beta \bigl( \O^{(i)} \bigr)$ and 
$B \in {\cal R}_\beta \bigl( \hat{\O}^{(i)} \bigr)' $,  
and
\begin{align*}( \eta_i \, , \, M N  \, \eta_i) &= \bigl(\Omega_\beta \, , \, M \Omega_\beta \bigr)
\bigl(\Omega_p^{(i)} \, , \, N \Omega_p^{(i)} \bigr)  
\\
( \chi_i \, , \, M N  \, \chi_i)  &= (\Omega_\beta \, , \, M \Omega_\beta)
(\Omega_\beta \, , \, N \Omega_\beta) 
\end{align*}
for $M \in {\cal M}_\beta \bigl( \O_\circ^{(i)} \bigr)$ and $N \in {\cal M}_\beta \bigl( \O^{(i)} \bigr)' $.

\vskip .5cm
So far there was no restriction on the relative size of the regions
$\O^{(i)}$ and $\hat {\O}^{(i)}$. We will now exploit this freedom: 
If the net of local observables $\O \to {\cal R}_\beta (\O)$ is regular from the outside, then
${\cal M}_\beta (\O) ' \cap {\cal M}_\beta (\hat{\O}) \to \C \cdot \unit$   
as $\hat{\O} \searrow \O$.   
Our aim is to control $\| \eta_i - \Omega_\beta \|$. 
The following lemma shows that in order to do so it is sufficient to control~$\| \chi_i - 
\Omega_\beta \|$. 

\begin{lemma}
{Let
$\bigl\{  ( \O_\circ^{(i)} , \O^{(i)} ) \bigr\}_{i \in \n}$ be a sequence of 
pairs of double cones. 
Then one can find a sequence of double cones $\{ \hat{\O}^{(i)} \}_{i \in \n}$ such that
(\ref{regions}) holds true and
$\lim_{i \to \infty} \| \chi_i - \eta_i \| =0$. 
}
\end{lemma}

\vskip .3cm

\Pr{Consider a sequences of pairs of double cones   
$\bigl\{ \bigl( \O_\circ^{(i)} , \O^{(i)} \bigr) \bigr\}_{ i \in \n} $  
eventually exhausting all of $\R^4$. For each $i \in \N$ fixed we consider a sequence 
of double cones $\bigl\{ \hat{\O}^{(i,k)} \bigr\}_{ k \in \n} $ such that
$\hat{\O}^{(i,k)} \searrow \O^{(i)}$ for $k \to \infty$. 
In order to ease the notation we set
\[  {\cal A}_i := {\cal M}_\beta \bigl( \O_\circ^{(i)} \bigr),
\qquad
{\cal B}_i := {\cal M}_\beta \bigl( {\cal O}^{(i)} \bigr),
\qquad 
{\cal C}_{i,k} := {\cal M}_\beta \bigl( \hat{\cal O}^{(i,k)} \bigr)  , \]
${\cal D}_{i} := {\cal A}_i \vee {\cal B}_i'$,
and 
${\cal E}_{i,k} := {\cal A}_i \vee {\cal C}_{i,k}' $. 
For each $i \in \N$ fixed, the sequence $\{ {\cal E}_{i,k}  \}_{k \in \n}$ of algebras 
satisfies 
${\cal E}_{i,k+1}  \subset {\cal E}_{i,k} $ (this follows from ${\cal C}_{i,k+1}  \subset {\cal C}_{i,k} $) and  
$\cap_k {\cal E}_{i,k} =  {\cal D}_i $.  
Now let~$\Omega_p^{(i,k)}$ denote the unique product vector in the natural positive 
cone ${\cal P}^\natural \bigl({\cal R}_\beta (\O^{(i)}) \vee {\cal R}_\beta (\hat {\O}^{(i,k)})', 
\Omega_\beta \bigr) $
satisfying
\[  \bigl(\Omega_p^{(i,k)} \, , \, AB   \Omega_p^{(i,k)} \bigr) =  (\Omega_\beta \, , \, A \Omega_\beta) 
(\Omega_\beta \, , \, B \Omega_\beta) \]
for all $A \in {\cal R}_\beta \bigl(\O^{(i)} \bigr)$ and $B \in {\cal R}_\beta \bigl(\hat {\O}^{(i,k)}\bigr)'$.
Note that
for $C_{i,k} \in {\cal C}_{i,k}'$
\[  \bigl(\Omega_p^{(i,k)} \, , \, C_{i,k} \Omega_p^{(i,k)} \bigr) 
= (\Omega_\beta \, , \, C_{i,k} \Omega_\beta) . \]
If we choose product vectors $\eta_{i,k}$  and $\chi_i$ 
in the natural cone ${\cal P}^\natural ({\cal D}_i, \Omega_\beta)$
such that
\[   ( \eta_{i,k} \, , \, MN \eta_{i,k}) = \bigl( \Omega_\beta \, , \, M \Omega_\beta \bigr)
\bigl( \Omega_p^{(i,k)} \, , \, N \Omega_p^{(i,k)} \bigr) \]
and
$( \chi_i \, , \, MN \chi_i) = (\Omega_\beta \, , \, M \Omega_\beta)
(\Omega_\beta \, , \, N \Omega_\beta)$
for all $M \in {\cal A}_i$ and $N \in {\cal B}_i'$, 
then by a result of Araki \cite{A 74} 
\[  \| \eta_{i,k} - \chi_i  \|^2 
\le \sup_{ D_i  \in {\cal D}_i , \| D_i \| = 1 } \bigl| (\eta_{i,k} , D_i \eta_{i,k} )  
- (\chi_i , D_i 
\chi_i) \bigr|    . \]
Now assume that for each $i \in \N$ fixed there exist a sequence $\{  E_{i,k} \in {\cal E}_{i,k} \mid 
\| E_{i,k} \| = 1 \}_{k \in \n}$ such 
that
\beq  \lim_{k \to \infty} \bigl| (\eta_{i,k} \, , \, E_{i,k} \eta_{i,k}  )  
- (\chi_i \, , \, E_{i,k} \chi_i) \bigr| \ge \epsilon_i . \label{contra} \eeq
We demonstrate that this leads to a contradiction. 
The linear functional $(\eta_{i,k} \, , \, \,. \, \eta_{i,k}  )  
- (\chi_i \, , \, \,. \, \chi_i)$ is ultra-weakly continuous on the von Neumann 
algebra ${\cal D}_i$. 
Therefore the sequence $\{   E_{i,k} \in {\cal E}_{i,k} \mid 
\| E_{i,k}  \| = 1 \}_{k \in \n}$ has a weak limit point
$w-\lim_{k \to \infty} E_{i,k}  =: D_i \in {\cal D}_{i} =
\cap_k {\cal E}_{i,k} $ 
such that 
\[   \bigl| (\eta_{i,k} , D_i \eta_{i,k} ) - (\chi_i , D_i \chi_i)  \bigr| 
> \frac{1 }{ 2 }
\epsilon_i
\qquad \forall k > k_i \] 
and some $k_i \in \N$, in contradiction to 
\[   \bigl| (\eta_{i,k} , E_{i,k} \eta_{i,k} ) - (\chi_i , E_{i,k} \chi_i)  \bigr| = 0 
\qquad \forall E_{i,k} \in {\cal E}_{i,k}, \quad \forall k \in \N . \] 
Therefore, the assumption (\ref{contra})
can not hold true.
It follows that there exists some $k_i \in \N$ such that 
\[ \sup_{D \in {\cal D}_{i} , \| D  \| = 1}
 \bigl| (\eta_{i, k} \, , \, D  \eta_{i,k}  )  
- (\chi_i \, , \, D  \chi_i)  \bigr|  <  \epsilon_i  
\qquad \forall k \ge k_i. \]
If we set $\hat{\O}^{(i)} : = \hat{\O}^{(i, k_i)}$, then we can choose $\epsilon_i$ such that
$\lim_{i \to \infty} \| \chi_{i} - \eta_i \| =0$.} 

We will now show that the
product vector $\chi$ converges to $\Omega_\beta$ if $\O_\circ $ and $\O$  tend to $\rr^4$
and the relative size of $\O_\circ $ and $\O$ obeys the restrictions imposed by the
cluster condition.

\begin{lemma}
{Let $\bigl\{ \bigl(\O_\circ^{(i)}, \O^{(i)} \bigr) \bigr\}_{i \in \nn} $ denote a sequence of pairs of double cones
with diameters~$\bigl(r_\circ^{(i)}, r^{(i)} \bigr)$, $i \in \N$.
Assume that $\lim_{i \to \infty} \bigl(r_\circ^{(i)} \bigr)^{d'} \bigl(\delta_\circ^{(i)}\bigr)^{-\gamma} = 0$.   
It follows that
$\| \chi_i - \Omega_\beta  \|  \to 0 $
as $i \to \infty$.}
\end{lemma}

\Pr{Since $\chi_{i}  \in {\cal P}^\natural
\bigl({\cal M}_\beta ( \O_\circ^{(i)}) \vee {\cal M}_\beta ( \O^{(i)})', \Omega_\beta \bigr)$,
we can again rely on the result of Araki~\cite{A 74} concerning the distance of two vectors
which belong to the natural positive cone ${\cal P}^\natural
\bigl({\cal M}_\beta ( \O_\circ^{(i)}) \vee {\cal M}_\beta ( \O^{(i)})', \Omega_\beta \bigr)$: 
\[ \| \chi_{i} - \Omega_\beta  \|^2 
\le \sup_{  
\| D_i \| = 1 } \bigl| (\chi_i , D_i \chi_i )  
- (\Omega_\beta , D_i 
\Omega_\beta) \bigr|    ; \]
where the supremum has to be evaluated over all elements
$D_i \in {\cal M}_\beta \bigl( \O_\circ^{(i)} \bigr) \vee {\cal M}_\beta \bigl( \O^{(i)}\bigr)' $. 
Thus~$\lim_{i \to \infty} \| \chi_i - \Omega_\beta \| = 0 $  
follows from the cluster condition (\ref{cluster}) and the assumptions conerning the relative size of
$\O_\circ$ and $\O$ stated in the lemma. }
 
Combining Lemma 4.1 and Lemma 4.2 we conclude that $\lim_{i \to \infty} \| \eta_i - \Omega_\beta \| = 0$
for an appropriate choice of the relative size of $\O_\circ^{(i)}$, $\O^{(i)}$ and $\hat \O^{(i)}$.

\vskip  1cm
\subsection{Bounds on the quasi-partition function}

\noindent
Let us consider  a sequence $\{ \Lambda_i \} = \bigl\{ \bigl( \O_\circ^{(i)} , \O^{(i)} , \hat{\O}^{(i)} \bigr) \bigr\}$ of 
triples of double cones with diameters~$(r^{(i)}_\circ , r^{(i)} , \hat{r}^{(i)})$. 
In order to ensure that 
(for $0 < \alpha < 1 /2$ and $n \in \N$ fixed) the
`quasi-partition function' 
\[  Z_{\Lambda_i}(\alpha, n) := \Tr \, \bigl( E_{\Lambda_i} 
\Delta_{p, i}^\alpha E_{\Lambda_i} \bigr)^n , \qquad \Lambda_i = 
\bigl( \O_\circ^{(i)}, \O^{(i)} , \hat{\O}^{(i)} \bigr),  \] 
is bounded from below as $i \to \infty$, it is necessary that $\O^{(i)}$ grows rapidely 
with $\O_\circ^{(i)}$. Otherwise the energy contained 
in the boundary, which is necessary to decouple the 
local region from the outside, lessens the eigenvalues
of $E_{\Lambda_i} \Delta_{p, i}^\alpha E_{\Lambda_i}$ so drastically that
it outruns the increase in the number of states 
contributing to the trace by enlarging~$\O_\circ^{(i)}$. 
Following once again \cite{BJu 89} we will now demonstrate that the condition on the 
relative size of $r_\circ^{(i)}$ and $r^{(i)}$
which we imposed in order to show that $\chi_i$ converges to $\Omega_\beta$ is already sufficient 
to exclude this possibility.

\begin{lemma}
{(Buchholz and Junglas). Let $\{ \Lambda_i \}_{i \in \n}$ be a 
sequence of triples of increasing space-time regions such that 
$\|  \eta_i - \Omega_\beta  \|  \to 0$ for $i \to \infty$.
It follows that
$\H_{\Lambda_i}$ tends to the whole Hilbert space $\H_\beta$, i.e., $s-\lim_{i \to \infty} 
E_{\Lambda_i} = \unit $. }
\end{lemma}

\vskip .3cm

\Pr{By assumption $\eta_i = W^*_i  \bigl(\Omega_\beta \otimes \Omega_p^{(i)} \bigr) $ 
converges to $\Omega_\beta$. Therefore
the unitary operators~$W_i$ specified in (\ref{pv}) fulfill
$W^*_i  \bigl(\Phi \otimes \Omega_p^{(i)} \bigr) \to \Phi $
for $\Phi \in \H_\beta $
as $i \to \infty$. Recall that $E_{\Lambda_i} = W_i^* \bigl(\unit \otimes P_{\Omega_p^{(i)}} \bigr)
W_i$, where $P_{\Omega_p^{(i)}}$ denotes the projection onto~$\C \cdot \Omega_p^{(i)}$.
Hence
\[ E_{\Lambda_i}  \Phi =
W^*_i  \bigl(\unit \otimes P_{\Omega_p^{(i)}} \bigr) W_i
\bigl( \Phi - W^*_i  (\Phi \otimes \Omega_p^{(i)}) \bigr)
+ W^*_i  \bigl( \Phi \otimes \Omega_p^{(i)} \bigr) \to \Phi 
\quad
\forall \Phi \in \H_\beta , \]
as $i \to \infty$. I.e.,  $s-\lim_{i \to \infty } E_{\Lambda_i}  = \unit $.}
\begin{lemma}
{Let $\bigl\{ \bigl(\O_\circ^{(i)}, \O^{(i)} \bigr) \bigr\}_{i \in \nn} $ denote a sequence of pairs of double cones
with diameters~$(r_\circ^{(i)}, r^{(i)} )$, $i \in \N$.
Assume that $\lim_{i \to \infty} \bigl(r_\circ^{(i)} \bigr)^{d'} \bigl(\delta_\circ^{(i)}\bigr)^{-\gamma} = 0$.   
It follows that
\[   \lim \inf_i \Tr \, \bigl( E_{\Lambda_i} 
\Delta_{p, i}^\alpha E_{\Lambda_i} \bigr)^n 
> 0 \qquad \forall n \in \N. \]
}
\end{lemma}

\vskip .3cm

\Pr{By definition, $\Delta_{p,i}^{1/2}$ is a positive operator. 
The vector $\Omega_p^{(i)}$ is the unique eigenvector of~$H_p$ for the simple eigenvalue
$\{ 0 \}$. Let $\{ \Omega_\beta, \Psi_1, \Psi_2, \ldots \}$ be an orthonormal basis in $\H_\beta$ and set $\Gamma_j^{(i)} = W_i^* \bigl( \Psi_j^{(i)}  \otimes \Omega_p \bigr) \in \H_\beta$. 
For $0< \alpha < 1/2$ and $j \in \nn$ this implies that~$\bigl( \Gamma_j^{(i)}  \, , \,
\Delta_{p,i}^{2\alpha} \Gamma_j^{(i)}  \bigr) = \bigl( \Psi_j \, , \,
\Delta^{2\alpha} \Psi_j \bigr) (\Omega_p \, , \, \Omega_p)> 0$.
Since  $s-\lim_{i \to \infty}
E_{\Lambda_i} = \unit $, it follows that 
\begin{align*}\lim \inf_i \Tr \, \bigl( E_{\Lambda_i} 
\Delta_{p, i}^\alpha E_{\Lambda_i} \bigr)^2 
&\ge \lim \inf_i \sum_{j=1}^\infty \Bigl( E_{\Lambda_i}  \Gamma_j^{(i)}  \, , \,
\Delta_{p, i}^\alpha  E_{\Lambda_i} 
\Delta_{p, i}^\alpha  E_{\Lambda_i}  \Gamma_j^{(i)}  \Bigr)
\\
&=\lim \inf_i \sum_{j=1}^\infty \Bigl( \Delta_{p, i}^\alpha   \Gamma_j^{(i)}  \, , \,
E_{\Lambda_i} 
\Delta_{p, i}^\alpha    \Gamma_j^{(i)}  \Bigr)
\\
&= \lim \inf_i \sum_{j=1}^\infty ( \Psi_j \, , \,
\Delta^{2\alpha} \Psi_j )
> 0 .
\end{align*}
}

\vskip  1cm

\subsection{Commutator estimates}

\noindent
The unit ball in $\A^*$ is weak$^*$-compact. Thus for every net of states
$\Lambda(\O_\circ, \O, \hat {\O}) \to \omega_\Lambda $
there exists  a subnet of  $\{ \omega_{\Lambda_i} \}_{i \in I}$ converging to 
some state $\omega$. Whether or not this state is a $(\tau, n \alpha \beta)$-KMS
state depends on the energy contained in the boundary, i.e.,
the choice of the relative size of~${\cal O}_\circ^{(i)}$,~$\O^{(i)}$ and~$ \hat {\O}^{(i)}$.
We show that the necessary quantitative
information restricting the surface energy can be drawn from the bounds on the 
nuclear norm of the map~$\Theta_{\alpha, {\cal O}}$ 
introduced in (\ref{nuc}) and the cluster condition (\ref{cluster}).

\vskip .5cm
Let 
$\bigl\{ \Lambda_i = \bigl( \O_\circ^{(i)} , \O^{(i)} , \hat{\O}^{(i)} \bigr)\bigr\}_{i \in \n}$ be a sequence\footnote{Note that it is sufficient
to work with sequences if the operators $a$ and $b$ appearing in (\ref{limit}) are fixed.}   of 
triples of double cones with diameters~$\bigl(r^{(i)}_\circ , r^{(i)} , \hat{r}^{(i)} \bigr)$, $i \in \N$. 
We will now exploit the fact that the elements of $\A_p$, $p \in \N$, introduced at 
the end of Subsection~2.2,
have good localization properties in space-time: we will show that 
there exists some $p \in \N$ such that
\beq \label{limit}   \bigl| \omega_{\Lambda_i} \bigl( a \tau_{i n\alpha \beta} (b) \bigr) -
\omega_{\Lambda_i} (b a ) \bigr| < \epsilon_i \qquad \forall a,b \in \A_p, \eeq 
where $\epsilon_i \searrow 0$ as $i \to \infty$. 
Thus the surface energy can be controlled 
by adjusting the relative size of~$r^{(i)}_\circ$, $r^{(i)}$  and $\hat{r}^{(i)}$.

Inspecting the definition (\ref{lKMS}) of $\omega_{\Lambda_i}$ 
we recognize that in order to prove (\ref{limit}) it is sufficient to control
\[  \Tr \, \,  \rho_{\Lambda_i} \,\,  \pi_\beta (a) 
\bigl[ \pi_\beta \bigl( \tau_{i k \alpha \beta} (b) \bigr) , E_{\Lambda_i} \bigr]  , \qquad
k = 1, \ldots, n. \] 
Let us consider the case $n=2$.
Let $a, b \in \A_p$, $p \in \N$ fixed.
It follows that
$\tau_{ik\alpha \beta} (b) \in \A_p $ for $k = 1, 2$. 
Since $a$ and $b$ as well as $ c:= \tau_{i\alpha \beta} (b)$ and $ d := \tau_{2i\alpha \beta} (b)$ 
are almost localized in $\O_\circ^{(i)}$ for $i$ sufficiently large,
they almost commute with $E_{\Lambda_i}$. 
For example,
\begin{align*} \Bigl|  \Tr \, \rho_{\Lambda_i} \, \pi_\beta (a) \bigl[ \pi_\beta \bigl( \tau_{i 2\alpha \beta} (b) \bigr) , 
E_{\Lambda_i} \bigr]  \Bigr|  &=
\frac{  \Bigl| \Tr \, [ \pi_\beta \bigl( \tau_{i 2\alpha \beta} (b) \bigr) , 
E_{\Lambda_i} ]   \cdot 
\bigl( \Delta_{p, i}^\alpha   
E_{\Lambda_i} \bigr)^2 \pi_\beta (a)  \Bigr|
}{ 
\Tr \, \bigl( \Delta_{p, i}^\alpha E_{\Lambda_i} \bigr)^2 } 
\\
& \le  \frac{ \| \,  [ \pi_\beta \bigl( \tau_{i 2\alpha \beta} (b) - d_i \bigr),  E_{\Lambda_i} ] \,  \|
\cdot
\Tr \, | \bigl( \Delta_{p, i}^\alpha 
E_{\Lambda_i} \bigr)^2| 
\cdot
\| a  \|  }{ \Tr \, 
\bigl( \Delta_{p, i}^\alpha E_{\Lambda_i} \bigr)^2} 
\\
& \le  \frac{ 2  \| a  \|  }{ \Tr \, \bigl( \Delta_{p, i}^\alpha E_{\Lambda_i} \bigr)^2} 
\,
\|  \tau_{i 2\alpha \beta} (b) - d_i \| 
\cdot
\bigl( \Tr \, | \Delta_{p, i}^\alpha E_{\Lambda_i} | \bigr)^2  .
\end{align*}
Here $d_i \in \A \bigl(\O_\circ^{(i)} \bigr)$ denotes a local approximation of 
$d:=\tau_{i 2\alpha \beta} (b) \in \A_p$ which satisfies~$[E_{\Lambda_i},d_i] = 0$. 
Thus
\beq
\Bigl| \Tr \, \rho_{\Lambda_i} \, \pi_\beta (a) \bigl[ \pi_\beta \bigl( \tau_{i 2\alpha \beta} (b) \bigr) , 
E_{\Lambda_i} \bigr]   \Bigr| \le \frac{c_1 }{ \Tr \, \bigl( \Delta_{p, i}^\alpha E_{\Lambda_i} \bigr)^2}
\cdot {\rm e}^{- c_2 (r^{(i)}_\circ)^{2p}} \cdot {\rm e}^{ c_3 (r^{(i)})^d}   
\label{com}
\eeq
for certain positive constants $c_1= 2 C \, \| a  \| $, $c_2 = \kappa / (2^{2p})$ and $c_3 = 2c \bigl(\alpha^{-m} + (1/2 -\alpha)^{-m} \bigr)$, where $m >0$.
In the last inequality we made use  of Proposition 3.3 and  
the second part of Lemma~2.1.  
Inspecting the r.h.s.\ of~(\ref{com}) closely, we find that the numerator  
vanishes in the limit~$i \to \infty$, 
if~$\exp \bigl(- c_2 (r^{(i)}_\circ)^{2p} \bigr) \cdot \exp \bigl(c_3 (r^{(i)})^d \bigr)$ goes to zero as $i \to \infty$.
As has been shown in the previous section,
the denominator does not vanish as $i \to \infty$, but is bounded from below by some 
positive constant, if $\lim_{i \to \infty}  (r_\circ^{(i)})^{d'} (\delta_\circ^{(i)})^{-\gamma} = 0$. 

In other words, the distance $\delta_\circ^{(i)}$ has to grow sufficiently fast such that 
$\eta_i \to \Omega_\beta$, and $p$ has to be chosen  
sufficiently large such that the elements in $\A_p$ are sufficiently well localized to
fulfill the boundary condition (\ref{almost}) up to some small error term. 

\medskip
We will now  
establish the KMS property for all weak limit points of~$\{ \omega_{\Lambda} \}$,
provided the regions $\Lambda_i = \bigl( \O_\circ^{(i)}, \O^{(i)}, \hat {\O}^{(i)} \bigr)$ 
tend to the whole space-time in 
agreement with the restrictions imposed on the relative size of 
$r_\circ^{(i)}$,  $r^{(i)}$ and $\hat{r}^{(i)}$.

\begin{theoreme}
{Assume that both the nuclearity condition (\ref{nuc}) and the cluster condition~(\ref{cluster}) hold. 
Then there exists a choice of triples of space-time regions $\Lambda_i$
such that
every weak limit point of the (generalized) sequence $\{ \omega_{\Lambda_i} \}_{i \in I}$ is a 
$\tau$-KMS state at temperature~$1 / \beta'> 0$. }
\end{theoreme}

\vskip .3cm

\Pr {Let $n \in \N$ and $0 < \alpha < 1/2$ be fixed such that $\beta'= n \alpha \beta$. Moreover, let
$\Lambda_i = \bigl( \O_\circ^{(i)}, \O^{(i)}, \hat {\O}^{(i)} \bigr)$ be a sequence of 
triples of double cones 
with diameters $r_\circ^{(i)}$, $r^{(i)}$ and $\hat{r}^{(i)}$ such that
$\lim_{i \to \infty} \bigl(r_\circ^{(i)} \bigr)^{d'} \bigl(\delta_\circ^{(i)}\bigr)^{-\gamma} = 0$
and 
$\hat{\O}^{(i)} \searrow \O^{(i)}$  
sufficiently fast  as $i \to \infty$ such that
$ \lim_{i \to \infty} \| \eta_{i} - \Omega_\beta \|  = 0 $.   
\vskip 0cm
Let us recall: the nuclearity condition fixes the constants $d$ and $m$ and the cluster condition fixes the constants
$d'$ and $\gamma$.
We will now fix $ p \in \nn$. Taking into account the restrictions on the relative size of $r^{(i)}_\circ$ and $r^{(i)}= r^{(i)}_\circ + 2 \delta_\circ^{(i)}$
imposed by the cluster condition (\ref{cluster})---it is sufficient that $\bigl(r_\circ^{(i)}\bigr)^{d'} 
\bigl(\delta_\circ^{(i)}\bigr)^{- \gamma}$
goes to zero as $i$ goes to infinity---we can now chose $p$ such that 
$\exp \bigl(- c_2 (r^{(i)}_\circ)^{2p} \bigr) \cdot \exp \bigl(c_3 (r^{(i)})^d \bigr)$ goes to zero as $i \to \infty$.
\vskip .2cm
\vskip .2cm
Let $a,b \in \A_p$ and consider the case $n= 2$.  
\vskip .2cm
\noindent
i.)  Let $\omega_{2 \alpha \beta}$ denote the limit state 
of a convergent subnet $\{ \omega_{\Lambda_i} \}_{i \in I}$.  
For every~$\epsilon > 0$ we can find an index $i \in I$ such that
\[  \bigl| \omega_{2 \alpha \beta} \bigl( a \tau_{i 2 \alpha \beta} (b) -ba \bigr) \bigr|  
 \le \bigl|   
\omega_{\Lambda_i}    \bigl( a \tau_{i 2 \alpha \beta} (b) -ba \bigr)  
\bigr| 
+ \epsilon . \]
\vskip .2cm
\noindent
ii.) We now approximate $\tau_{i 2 \alpha \beta} (b) , \tau_{i \alpha \beta} (b)$ and
$b$ by local elements in $\A \bigl( \O_\circ^{(i)} \bigr)$ and
apply the  commutator estimate (\ref{com}) several times: for
suitable (large) $i \in \nn$ we find
\begin{align*}
\bigl|  \omega_{2 \alpha \beta}  \bigl( a \tau_{i 2 \alpha \beta} (b) -ba \bigr)  \bigr| 
& \le 
\Bigl| \frac{  
\Tr \, \pi_\beta \bigl( a \tau_{i 2 \alpha \beta} (b)\bigr)    
\bigl( E_{\Lambda_i} \Delta_{p, i}^\alpha E_{\Lambda_i} \bigr)^2
}{ \Tr \, \bigl(  \Delta_{p, i}^\alpha E_{\Lambda_i} \bigr)^2} - \omega_{\Lambda_i}  (ba) \Bigr| 
+ \epsilon 
\\
& \le \Bigl|  \frac{  
\Tr \, \pi_\beta (a) E_{\Lambda_i} \pi_\beta \bigl(\tau_{i 2 \alpha \beta} (b)\bigr)    
\bigl( \Delta_{p, i}^\alpha E_{\Lambda_i} \bigl)^2 }{ \Tr \, \bigl(  
\Delta_{p, i}^\alpha E_{\Lambda_i} \bigr)^2}
- \omega_{\Lambda_i}  (ba) \Bigr| 
+ 2 \epsilon \\
& =  \Bigl| \frac{  
\Tr \, \pi_\beta (a) E_{\Lambda_i} \Delta_{p, i}^\alpha 
\pi_\beta \bigl(\tau_{i \alpha \beta} (b)\bigr)    
 E_{\Lambda_i}\Delta_{p, i}^\alpha
E_{\Lambda_i} }{ \Tr \, \bigl(   
\Delta_{p, i}^\alpha E_{\Lambda_i} \bigr)^2}
- \omega_{\Lambda_i}  (ba) \Bigr| 
+ 2 \epsilon 
\\
& \le  \Bigl| \frac{ 
\Tr \, \pi_\beta (a) \bigl( E_{\Lambda_i} \Delta_{p, i}^\alpha \bigr)^2
\pi_\beta (b) E_{\Lambda_i} }{ \Tr \, \bigl(  
\Delta_{p, i}^\alpha E_{\Lambda_i} \bigr)^2}
- \omega_{\Lambda_i}  (ba) \Bigr| 
+ 3 \epsilon \\
& \le  \Bigl| \frac{  
\Tr \, \pi_\beta (a) \bigl( E_{\Lambda_i} \Delta_{p, i}^\alpha
E_{\Lambda_i} \bigr)^2 \pi_\beta (b)
}{ \Tr \, \bigl(   
\Delta_{p, i}^\alpha E_{\Lambda_i} \bigr)^2 }
- \omega_{\Lambda_i}  (ba) \Bigr| 
+ 4 \epsilon 
\\
& = 4 \epsilon .
\end{align*}
Thus $\omega_{2 \alpha \beta} \bigl (a \tau_{2 i \alpha \beta}(b) \bigr) 
= \omega_{2 \alpha \beta} (b a) $ 
for all $a, b \in \A_p $.  Now recall that $\A_p$ (for each $p \in \N$) is a
$\tau$-invariant $\ast$-subalgebra of the set $\A_\tau$ of analytic elements of~$\A$ with respect to~$\tau$.
Consequently, $\omega_{2 \alpha \beta}~$ is a $(\tau, 2 \alpha \beta)$-KMS state.
\vskip .0cm
Similar results for arbitrary $n \in \N$ can be established by the same line of arguments
but with considerable more effort.}

Once we have constructed a $(\tau, \beta')$-KMS state $\omega_{\beta'}$,
the GNS-representation $\pi_{\beta'}$ leads to a new  
thermal field theory
\[  \O \to {\cal R}_{\beta'} (\O) := \pi_{\beta'} \bigl( \A(\O) \bigr)'' , 
\qquad \O \in \R^4, \]
acting on a new Hilbert space
$\H_{\beta'}$ with GNS-vector $\Omega_{\beta'}$. If $\beta \ne \beta'$, then
the new thermal field theory will not
be unitarily equivalent to the old one \cite{T}. In fact,
there might even be several extremal $(\tau, n \alpha \beta)$-KMS states,
which induce unitarily inequivalent representations, i.e., ``disjoint
thermal field theories'', at the same temperature $1 / \beta' = (n \alpha \beta)^{-1}$.

\vskip .5cm

\noindent
{\it  Acknowledgements.\/} The present work started in collaboration with D.\ Buchholz. 
The final formulation is strongly influenced by his constructive criticism and
by several substantial hints. Kind hospitality of the 
II.\ Institute for theoretical physics, University of Hamburg,
the Institute for theoretical physics, University of Vienna, the 
Erwin Schr\"odinger Institute (ESI), Vienna, and the Dipartimento di 
Matematica, Universita di Roma ``Tor Vergata" is gratefully acknowleged. 
This work was financed by 
the Fond zur F\"orderung der Wissenschaft\-lichen 
Forschung in Austria, Proj.\ Nr.\ P10629 PHY and 
a fellowship of the Operator Algebras Network, EC TMR-Programme.

\end{document}